\documentclass[11pt,a4paper]{article}
\pdfoutput=1
\usepackage{jheppub}
\usepackage{amsmath}
\usepackage{epsfig}
\usepackage{amssymb}
\usepackage{graphics}
\usepackage[active]{srcltx}
\usepackage{epstopdf}
\usepackage{pdfsync}
\usepackage{shuffle}

\newcommand{\bea}{\begin{eqnarray}}
\newcommand{\eea}{\end{eqnarray}}
\newcommand{\bean}{\begin{eqnarray*}}
\newcommand{\eean}{\end{eqnarray*}}
\newcommand{\nn}{\nonumber \\}

\def\W #1{\widetilde{#1}}
\def\WH #1{\widehat{#1}}

\def\Tr{\mathop{\rm Tr}}

\def\eref#1{(\ref{#1})}

\def\a{{\alpha}}

\def\b{{\beta}}

\def\eps{\epsilon}

\def\Sl{\sum\limits}

\def\Label#1{\label{#1}%
  \smash{\hbox to0pt{\raise1ex\hbox{\tiny[#1]}\hss}}}

\def\cyclic#1{\mbox{Cyclic}\{#1\}}

\def\YMs{{\tiny\mbox{YMs}}}
\def\CHY{{\tiny \mbox{CHY}}}
\def\PT{{\mbox{PT}}}
\def\EYM{{\tiny \mbox{EYM}}}
\def\Pf{{\mbox{Pf}}}

\def\spaa #1{\langle #1\rangle}

\title{Understanding the Cancelation of Double Poles in the Pfaffian of CHY-formulism}
\author[a]{Rijun Huang,}
\author[b]{Yi-Jian Du\footnote{The correspondence author.},}
\author[c,d]{and Bo Feng \footnote{The unusual ordering of authors instead of the
standard alphabet ordering is for young researchers to get proper recognition of
contributions under the current out-dated practice in China. }}

\affiliation[a]{Department of Physics and Institute of Theoretical Physics, Nanjing Normal University,\\
 No.1 Wenyuan Road, Nanjing 210046, P. R. China.}

\affiliation[b]{Center for Theoretical Physics, School of Physics and Technology,
Wuhan University, \\
No.299 Bayi Road, Wuhan 430072, P.R. China.}
\affiliation[c]{Zhejiang Institute of Modern Physics, Department of Physics,
 Zhejiang University,\\
 No.38 Zheda Road, Hangzhou 310027, P.R. China.}
\affiliation[d]{Center of Mathematical Science,
  Zhejiang University,\\
  No.38 Zheda Road, Hangzhou 310027, P.R. China.}

\emailAdd{huang@nbi.dk}
\emailAdd{yijian.du@whu.edu.cn}
\emailAdd{fengbo@zju.edu.cn}

\date{\today}
\abstract{For a physical field theory, the tree-level amplitudes should possess only single poles.
However, when computing amplitudes with Cachazo-He-Yuan (CHY) formulation, individual terms in the intermediate
steps will contribute higher-order poles. In this paper, we investigate the cancelation of higher-order
poles in CHY formula with Pfaffian as the building block. We develop a diagrammatic rule for
expanding the reduced Pfaffian. Then by organizing diagrams in appropriate groups and applying
the cross-ratio identities, we show that all potential contributions to higher-order poles in
the reduced Pfaffian are canceled out, i.e., only single poles survive
in Yang-Mills theory and gravity. Furthermore, we show the cancelations of higher-order
poles in other field theories by introducing appropriate truncations, based on the single pole structure of
Pfaffian.}
\keywords{Scattering Amplitude, CHY-Formulation, Integration Rules}

\begin{document}
\maketitle \flushbottom

\section{Introduction}
\label{secIntroduction}

The Cachazo-He-Yuan(CHY) formula
\cite{Cachazo:2013gna,Cachazo:2013hca, Cachazo:2013iea,
Cachazo:2014nsa,Cachazo:2014xea} provides a new perspective to
understand scattering amplitudes for massless particles in arbitrary
dimensions. The skeleton of CHY-formula consists of  so-called
scattering equations
\bea  \mathcal{E}_a=\Sl_{ b\neq a,b=1}^n{s_{ab}\over
{z_a-z_b}}=0~~~,~~~a=1,\ldots,n~,~~~\label{eq:ScatteringEquations}
\eea
where $z_a$'s, $a=1,\ldots,n$ are complex variables. The Mandelstam
variable  $s_{ab}$ is defined by $s_{ab}=2k_a\cdot k_b$ and $k_a$
denotes the momentum of external particle $a$. M{\"o}bius invariance
of scattering equations allows us to reduce the number of
independent equations to $(n-3)$, while the equations
\eqref{eq:ScatteringEquations} have $(n-3)!$ independent solutions.
Based on the scattering equations \eqref{eq:ScatteringEquations},
CHY formula expresses an $n$-point tree-level scattering amplitude
$A_n$ for massless particles as follows,
\bea A_n=\int\frac{dz_{1}\ldots
dz_{n}}{\text{Vol}\left[SL(2,\mathbb{C})\right]}{\prod_{a}}\,'\delta
\left(\sum_{b\neq
a}\frac{s_{ab}}{z_{ab}}\right)\mathcal{I}^{\CHY}_{n}~.~~~
\label{eq:CHY} \eea
The scattering equations and the measure of CHY-integrals are
universal for all theories while $\mathcal{I}_n^{\CHY}$ encodes all
the information, including external polarizations, for a specific
field theory.

Although the CHY-formulation is simple and beautiful, the evaluation
of amplitude is difficult. It is almost impossible to perform
computation beyond five points by directly solving the scattering
equations due to the complicated nature of algebraic system. A few
studies on the direct solutions can be found in
\cite{Kalousios:2013eca,Weinzierl:2014vwa,Lam:2014tga,Du:2016blz,Du:2016wkt,Du:2016fwe,He:2016vfi,He:2016dol,He:2016iqi,Zhang:2016rzb},
but restricted to four-dimension and at special kinematics.
Alternative methods for evaluating CHY-integrals without explicitly
solving scattering equations are proposed by several groups from
different approaches. Some of the methods borrow the ideas from
computational algebraic geometry, by use of Vieta formula
\cite{Kalousios:2015fya}, elimination theory
\cite{Cardona:2015eba,Cardona:2015ouc,Dolan:2015iln}, companion
matrix \cite{Huang:2015yka}, Bezoutian matrix
\cite{Sogaard:2015dba}. Based on the polynomial form
\cite{Dolan:2014ega} of scattering equations, polynomial reduction
techniques are also introduced in this problem
\cite{Bosma:2016ttj,Zlotnikov:2016wtk}. In
\cite{Chen:2016fgi,Chen:2017edo}, differential operators are applied
to the evaluation of CHY-integrals. More computational efficient
methods are also developed recently. Techniques for contour
integration of the CHY-integrals are proposed and sharpened in
\cite{Cachazo:2015nwa,Gomez:2016bmv,Cardona:2016bpi}, where the
concept of $\Lambda$ scattering equations is introduced, and it
results to a recursive computation of CHY-integrand from lower-point
sub-CHY-integrands until end up with some simple building blocks. In
\cite{Lam:2015sqb,Lam:2016tlk}, Feynman-like diagrams are introduced
for the evaluation of CHY-integral by some kind of rules, and in
\cite{Mafra:2016ltu}, Berends-Giele recursions also also applied to
the situations where CHY-integrands are products of two Parke-Taylor
factors. While along the other routine, string theory inspired
method has been developed systematically in
\cite{Baadsgaard:2015voa,Baadsgaard:2015ifa,Baadsgaard:2015hia} and
\cite{Huang:2016zzb}, named integration rule method. The discovery
of cross-ratio identity further sharpens the computational power of
integration rule method
\cite{Bjerrum-Bohr:2016juj,Cardona:2016gon,Bjerrum-Bohr:2016axv},
making it simple and automatic for evaluating any generic
CHY-integrand\footnote{By private communication, Yong Zhang provides a mathematica code for CHY evaluation. Readers who are interested in the code can contact the email address yongzhang@itp.ac.cn.}.

The integration rule method combined with cross-ratio identity has
now become an ideal tool for evaluating CHY-integrals. The
evaluation of amplitudes in the CHY-formulism can be performed with
only the knowledge of CHY-integrands, ignoring the solutions of
scattering equations, the CHY-integral measure, etc. However, some
issues do need further clarification. For the integration rule
method to be valid, the terms to be evaluated should be M\"obius
invariant. In the current case, it means the CHY-integrand behaves
as ${1\over z_i^4}$ under $z_i\to \infty$. While all the
CHY-integrands for the known field theories so far are by
construction M\"obius invariant, each single term in the expansion
of CHY-integrands is not apparently M\"obius invariant, which causes
trouble for the intermediate computation. This issue is not yet
completely solved, but for most CHY-integrands containing the
reduced Pfaffian of matrix $\Psi$, a rewriting of certain entries of
matrix $\Psi$ would be suffice to make every term in the expansion
of CHY-integrands M\"obius invariant. For the situations where
integration rule method is applicable, we then confront the double
pole (or more generically, higher-order pole) problem. For field
theories, the physical amplitudes should possess only single poles.
However in the setup of CHY-framework, terms in the expansion of
CHY-integrands would be evaluated to results of higher-order poles
in the intermediate steps. Of course summing over all results the
higher-order poles should be canceled by factors in the numerator,
but in most computations we would get a large size of data which
makes it impossible to simplify further in a normal desktop. Hence
the cancelation of higher-order poles is inexplicit in the final
result generated by integration rule method. It is not unexpected
that, the origin of higher-order poles can be traced back to the
CHY-integrand level, and a thorough understanding of how the
cancelation works out in the CHY-integrand level would be a crucial
step towards the generalization of CHY-formalism.

In this paper, we systematically study the cancelation of potential
higher-order poles in various field theories described by
CHY-integrands. This paper is organized as follows. In
\S\ref{secReview}, we provide a review on the CHY-integrands in
various field theories, the expansion of Pfaffian and cross-ratio
identities. A diagrammatical expansion of reduced Pfaffian is
provided in \S\ref{secRule}. The cancelation of double poles in
Yang-Mills theory and gravity are investigated in
\S\ref{secCancelation}, where explicit examples are provided.
General discussions on the cancelation of double poles for other
field theories are given in \S\ref{secTrunction}. Conclusion can be
found in \S\ref{secConclusion}, and in appendix we give detailed
studies on the off-shell and on-shell identities of CHY-integrands
and illustrate their applications to simplify complicated
CHY-integrands.

\section{A review of CHY-integrand, the expansion of Pfaffian
and cross-ratio identity}
\label{secReview}

In this section, we provide a review on the CHY-integrand of various
field theories and the related knowledge, e.g., the expansion of
Pfaffian, the cross-ratio identity and integration rules, which is
useful for later discussions.

\noindent {\bf The CHY-integrands:} the field theory is fully
described by its corresponding CHY-integrand $\mathcal{I}^{\CHY}$,
and in the concern of integration rule method, only CHY-integrand is
necessary for the evaluation of amplitude. The building block of
CHY-integrands are Parke-Taylor(PT) factor
\bea \PT(\alpha):={1\over
z_{\alpha_1\alpha_2}z_{\alpha_2\alpha_3}\cdots
z_{\alpha_{n-1}\alpha_n}z_{\alpha_n\alpha_1}}~~,~~z_{ij}=z_i-z_j~,~~~\label{PTfactor}\eea
and the Pfaffian and reduced Pfaffian of certain matrix. For
$n$-particle scattering, let us define the following three $n\times
n$ matrices $A,B,C$ with entries
\bea
\begin{array}{llll}
  A_{a\neq b}={k_a\cdot k_b\over z_{ab}}~,~~~&B_{a\neq b}={\epsilon_a\cdot \epsilon_b\over z_{ab}}~,~~~&C_{a\neq b}={\epsilon_a\cdot k_b\over z_{ab}}~,~~~&X_{a\neq b}={1\over z_{ab}}~,~~~ \\
A_{a=b}=0~,~~~&B_{a=b}=0~,~~~&C_{a=b}=-\sum_{c\neq a}{\epsilon_a\cdot k_c\over z_{ac}}~,~~~&X_{a=b}=0~.~~~
\end{array}      \label{ABCXmatrix}\eea
Special attention should be paid to the diagonal entries of matrix
$C$ since they will break the M\"obius invariance of terms in the
expansion of CHY-integrands. In the practical computation, the
definition
\bea C_{aa}=\sum_{i\neq a,t}(\epsilon_a\cdot k_{i}){z_{it}\over
z_{ia}z_{at}}~~~~\label{Caa}\eea
is adopted, which is equivalent to the original definition by
momentum conservation and scattering equations. This definition
provides a better M\"obius covariant representation, i.e., it is
uniform weight-2 for $z_a$ and weight-0 for others. The $z_t$ is a
gauge choice and can be chosen arbitrary. With matrices $A,B,C$, we
can define a $2n\times 2n$ matrix $\Psi$,
\bea \Psi=\left(
            \begin{array}{cc}
              A &~ -C^{T} \\
              C &~ B \\
            \end{array}
          \right)~,~~~\label{Psimatrix}
\eea
where $C^T$ is the transpose of matrix $C$.

With these building blocks (\ref{PTfactor}), (\ref{ABCXmatrix}) and
(\ref{Psimatrix}), we are able to construct CHY-integrands for a
great number of theories. For such purpose, the Pfaffian of
skew-symmetric matrix is introduced. The determinant of an
anti-symmetric matrix $\Psi$ is a perfect square of some polynomial,
and the Pfaffian $\Pf~\Psi$ is defined as the square root of the
determinant. In the solution of scattering equations, the $2n\times
2n$ matrix $\Psi$ is degenerate, so we need further to introduce the
reduced Pfaffian $\Pf~'~\Psi$ defined as
\begin{equation}
\Pf~'\Psi:=\frac{2 (-)^{i+j}}{z_{ij}}
\Pf~\Psi_{ (ij)}^{ (ij)}~,~~~\label{eq:rf}
\end{equation}
where $\Psi^{(ij)}_{(ij)}$ stands for the matrix $\Psi$ with the
$i$-th and $j$-th column and rows removed. Of course the definition
of Pfaffian and reduced Pfaffian applies to any skew-symmetric
matrices, for instance the matrix $A$ defined in (\ref{ABCXmatrix}).

With above definitions, we list the CHY-integrand for various
theories \cite{Cachazo:2016njl} as,
\begin{center} \begin{tabular}{|c|c|c|}
  \hline
  {\bf The described theory} & $\mathcal{I}_L$ & $\mathcal{I}_R$  \\\hline
 Bi-adjoint scalar & $\PT_n(\alpha)$ & $\PT_n(\beta)$ \\\hline
 Yang-Mills theory & $\PT_n(\alpha)$ & $\Pf~'\Psi_n$ \\\hline
  Einstein gravity & $\Pf~'\Psi_n$  & $\Pf~'\W\Psi_n $ \\\hline
  Einstein-Yang-Mills theory (single trace) & $\PT_m(\beta)~\Pf~\Psi_{n-m}$  & $\Pf~'\W\Psi_{n} $ \\\hline
  Born-Infeld theory & $(\Pf~'A_n)^2$ & $\Pf~'\Psi_n$ \\\hline
  Nonlinear sigma model & $\PT_n(\alpha)$ & $(\Pf~'A_n)^2$ \\\hline
  Yang-Mills-scalar theory & $\PT_n(\alpha)$ & $(\Pf~X_n)~(\Pf~'A_n)$ \\\hline
  Einstein-Maxwell-scalar theory & $(\Pf~X_n)~(\Pf~'A_n)$ & $(\Pf~X_n)(\Pf~'A_n) $\\\hline
  Dirac-Born-Infeld theory &$(\Pf~'A_n)^2$ & $(\Pf~X_n)~(\Pf~'A_n)$ \\\hline
  Special Galileon theory & $(\Pf~'A_n)^2$ & $(\Pf~'A_n)^2$\\
  \hline
\end{tabular}
\end{center}
where we have used the fact that the CHY-integrands
$\mathcal{I}^{\CHY}$ is a weight-4 rational functions of $z_i$'s
which can usually be factorized as product of two weight-2
ingredients $\mathcal{I}^{\CHY}=\mathcal{I}_L\times \mathcal{I}_R$.

\noindent{\bf The expansion of CHY-integrand:} the difficulty of
evaluation comes from the terms of Pfaffian in the CHY-integrands,
which would produce higher-order poles. So a genuine expansion of
Pfaffian is possible to simplify our discussion. In
\cite{Lam:2016tlk}, it is pointed out that the reduced Pfaffian
$\Pf~'\Psi$ can be expanded into cycles as,
\bea \Pf~'\Psi=-2^{n-3}\Sl_{p\in S_n}(-1)^p{W_IU_J\cdots U_K\over
z_I z_J\cdots z_K}~,~~~\label{eq:ExpandPf} \eea
where the permutation $p$ has been written into the cycle form with
cycles $I,J,\ldots,K$. The $z_I$ for a given cycle
$I=(i_1,i_2,\cdots,i_m)$ is defined as $z_{i_1i_2}z_{i_2i_3}\cdots
z_{i_mi_i}$. For length-$m$ cycle $I$, a constant factor
$(-1)^{m+1}$ should be considered, which sums together to give the
$(-)^p$ factor in (\ref{eq:ExpandPf}). The open cycle $W$ is defined
as
\bea W_I=\epsilon_{i\lambda}\cdot\left(F_{i_2}\cdot F_{i_3}\cdots
F_{i_{m-1}}\right) \cdot\epsilon_{j\nu}~,~~~\label{eq:W-cycle} \eea
in which $\epsilon_{i\lambda}$ and $\epsilon_{j\nu}$ denote the
polarizations   of particles $i,j$ respecting to the deleted rows
and columns (i.e., the gauge choice). The closed cycle $U$ is
defined as
\bea
\begin{array}{l}
U_I={1\over 2}\Tr(F_{i_1}\cdot F_{i_2}\cdots
F_{i_m})~,~~~~~~(\text{for $I$ contains more than one
element})~,~~~\\
 U_I=C_{ii}~~~~~~~~~~~~~~~~~~~~~~~~~~~~~~~~
 (\text{for $I$ contains only one element $i$})~.~~~
\end{array}~~~~\label{eq:U-cycle} \eea
In eqs. (\ref{eq:W-cycle}) and (\ref{eq:U-cycle}), $F^{\mu\nu}_{a}$
is defined as
\bea F_a:=k_a^{\mu}\epsilon_a^{\nu}- \epsilon_a^{\mu}
k_a^{\nu}~.~~~\eea

The Pfaffian which is used in e.g., EYM theory also have a similar
expansion,
\bea \Pf~\Psi_m=(-1)^{{1\over 2}m(m+1)}\Sl_{p\in
S_m}(-1)^p{U_IU_J\cdots U_K\over z_Iz_J\cdots
z_K}~,~~~\label{eq:ExpandPf2} \eea
where $\Psi_m$ is an $2m\times 2m$ sub-matrix of $\Psi_n$ by
deleting the rows and columns corresponding to a set of $(n-m)$
external particles.

For presentation purpose, we would use the following notation for
closed and open cycles,
\bea [a_1,a_2,\cdots,a_n]:=z_{a_1a_2}z_{a_2a_3}\cdots
z_{a_{n-1}a_{n}}~~~,~~~\langle
a_1,a_2,\cdots,a_n\rangle:=z_{a_1a_2}z_{a_2a_3}\cdots
z_{a_{n-1}a_{n}}z_{a_{n}a_1}~.~~~\label{notation}\eea

\noindent{\bf The cross-ratio identity and others:} to expand the
terms of Pfaffian with higher-order poles into terms with single
poles, we shall apply various identities on the CHY-integrands
\cite{Bjerrum-Bohr:2016juj,Cardona:2016gon,Bjerrum-Bohr:2016axv}.
Some identities are algebraic, for instance
\bea  {z_{ab} z_{dc}\over z_{ac} z_{bc}} ={ z_{ad}\over
z_{ac}}-{z_{bd}\over z_{bc}}~,~~~\label{z-identity}\eea
which does not require the $z$ to be the solutions of scattering
equations. We will call them {\sl off-shell identities}. The other
identities are valid only on the solutions of scattering equations,
and we will call them {\sl on-shell identities}. An important
on-shell identity is the cross-ratio identity,
\bea -1=\sum_{i\in A/\{a\},j\in A^c/\{b\}}{s_{ij}\over
s_A}{z_{ia}z_{jb}\over z_{ij}z_{ab}}~,~~~\label{crossratio} \eea
where $A$ is a subset of $\{1,2,...,n\}$ and $A^c$ is its complement
subset. Because of momentum conservation we have $s_A=s_{A^c}$. The
choice of $(a,b)$ is called the gauge choice of cross-ratio
identity, and different gauge choice will end up with different but
equivalent explicit expressions.

In the Appendix \ref{appendixIdentity} we will give detailed studies
on the various identities and their applications to reduce
complicated CHY-integrands to simple ones.

\noindent {\bf The order of poles:} during the process of
evaluation, the CHY-integrand is expanded into many M\"obius
invariant terms, with the generic form,
\bea {f(\epsilon,k)\over{\prod_{1\leq i< j\leq n}
z_{ij}^{\alpha_{ij}}}}~,~~~\label{termMobius} \eea
where $f(\epsilon,k)$ is kinematic factors, which is irrelevant for
the evaluation. The integration rule method provides a way of
examining the poles that to appear in the final result after
evaluation as well as the order of poles. The M\"obius invariant
term (\ref{termMobius}) can be represented by {\sl 4-regular graph},
where each $z_i$ is a node and a factor $z_{ij}$ in denominator is
represented by a solid line from $z_i$ to $z_j$ while a factor
$z_{ij}$ in numerator is represented by a dashed line. We would
generically express the factor $z_{ij}$ in numerator as $z_{ij}$
with negative $\alpha_{ij}$. In this setup, the possible poles of a
term (\ref{termMobius}) is characterized by the {\bf pole index
$\chi(A)$} \cite{Baadsgaard:2015voa,
Baadsgaard:2015ifa,Baadsgaard:2015hia}:
\bea \chi(A):=\mathbb{L}[A]-2(|A|-1)~.~~~\label{order-poles} \eea
Here, the linking number $\mathbb{L}[A]$ is defined as the number of
solid lines minus the number of dashed lines connecting the nodes
inside set $A$ and $|A|$ is the length of set $A$. For a set
$A=\{a_1,a_2,\ldots, a_m\}$ with pole index $\chi(A)$, the pole
behaves as ${1/\left(s_A\right)^{\chi(A)+1}}$ in the final result.
If $\chi(A)<0$, $s_A$ will not be a pole, while if $\chi(A)=0$,
$s_A$ will appear as a single pole, and if $\chi(A)>0$, it will
contributes to higher-order poles. The higher-order poles do appear
term by term in the expansion of CHY-integrals. For example, in
Yang-Mills theory with a single reduced Pfaffian, we can have double
poles in some terms. While in Gravity theory with two reduced
Pfaffian, we can have triple poles in some terms.

As mentioned, the wight-4 CHY-integrand $\mathcal{I}^{\CHY}$ has a
factorization $\mathcal{I}^{\CHY}=\mathcal{I}_L\times \mathcal{I}_R$
where $\mathcal{I}_L,\mathcal{I}_R$ are weight-2 objects. We can
also define the pole index for them as
\bea \chi_L(A):=\mathbb{L}[A]_{{\cal
I}_L}-(|A|-1)~~~,~~~\chi_R(A):=\mathbb{L}[A]_{{\cal I}_R}-(|A|-1)
~,~~~\label{order-poles-LR} \eea
and
\bea  \chi(A)=\chi_L(A)+\chi_R(A)~,~~~\label{order-poles-new}\eea
where the linking number is now counted inside each $\mathcal{I}_L$
or $\mathcal{I}_R$. It is easy to see that, for PT-factor given in
(\ref{PTfactor}) we will always have $\chi(A)\leq 0$. For the
reduced Pfaffian or Pfaffian of sub-matrix given in
\eqref{eq:ExpandPf} and \eqref{eq:ExpandPf2}, we have $\chi(A)\leq
1$. The condition $\chi(A)= 1$ happens when and only when the set
$A$ contains one or more cycles (i.e., a cycle belongs to $A$ or
their intersection is empty). This explains the observation
mentioned above that, for CHY-integrands given by the product of
PT-factor and reduced Pfaffian, individual terms can contribute to
double poles, while for gravity  theory with CHY-integrands given by
the product of two reduced Pfaffian, individual terms can contribute
to triple poles.

\section{Diagrammatic rules for the expansion of Pfaffian}
\label{secRule}

To evaluate amplitudes via CHY-formula, we should expand the
(reduced) Pfaffian as shown in \eqref{eq:rf} and
\eqref{eq:ExpandPf2}. In this expansion, there  are two information.
One is the variables $z_i$'s and the other one is the kinematics
$(k_i$'s, $\eps_i$'s). The $W_I, U_I$ factors given in
\eqref{eq:W-cycle},\eqref{eq:U-cycle} are compact collection of many
terms and since each term has its individual character, further
expansion of $W$ and $U$-cycles into terms of products of $(k_i\cdot
k_j)$, $(\epsilon_i\cdot k_j)$ and $(\epsilon_i\cdot\epsilon_j)$ is
needed. In this section, we establish a diagrammatic rule for
representing this expansion.

\subsection{Rearranging the expansion of Pfaffian}

In \eqref{eq:ExpandPf} and \eqref{eq:ExpandPf2} we sum over all
possible permutations $p$ of $n$ elements. This sum can be
rearranged as follows. We sum over the distributions of $n$ elements
into possible subsets and then sum over all permutations for each
subset in a given distribution. Then, for any given term $\cdots U_I
\cdots$ containing a cycle $U_I={1\over 2}\Tr{Tr}(F_{i_1}\cdot
F_{i_2}\cdots F_{i_m})$ ($m>2$), we can always find another term
which only differs from the former one by reflecting the
$U_I$-cycle. For example, for $n=4$, we can have a $(1)(234)$ and
also a $(1)(432)$ which are related by a refection of the second
cycle $(234)$. Since both the $U$-cycle and PT-factor satisfy the
same reflection relation,
\bea &&U_I:= {1\over 2}\text{Tr}(F_{i_1}F_{i_2}\cdots
F_{i_m})=(-1)^m{1\over 2} \text{Tr}(F_{i_1}F_{i_2}\cdots F_{i_m}):=
(-1)^m U_{I'}~,~~~\nn
&&z_I:= {1\over z_{i_1i_2}z_{i_2i_3}\cdots z_{i_mi_1}}
=(-1)^m{1\over z_{i_1i_m}\cdots z_{i_3i_2}z_{i_2i_1}}:= (-1)^m
z_{I'}~,~~~\nonumber \eea
we can pair them together as
\bea \left[(-1)^I{U_I\over z_I}+(-1)^{I'}{U_{I'}\over
z_{I'}}\right](-1)^J\cdots (-1)^K{U_J\cdots U_K\over z_J\cdots
z_K}=(-1)^{I+J+\cdots +K}{\W {U_I}U_J\cdots U_K\over z_Iz_J\cdots
z_K}~,~~~ \eea
where the sign $(-1)^I$ is $1$ when $I$ has odd number of elements
and $(-1)$ when  $I$ has even number of elements. The $\W U_I$ is
defined as
\bea \W {U_I}:=2U_I={\Tr}(F_{i_1}F_{i_2}\cdots
F_{i_m})~~~,~~~m>2~.~~~\label{WU-def} \eea
The cases with $m=1$ and $m=2$ are not included since the refections
of cycles $(i_1)$ and $(i_1i_2)$ are themselves. So we define $\W U=
U$ for $m=1,2$. The $W$-cycle is not included since its two ends
have been fixed. Using this manipulation, we rewrite the expansion
of reduced Pfaffian and Pfaffian of sub-matrix \eqref{eq:ExpandPf},
(\ref{eq:ExpandPf2}) as
\bea \Pf~'\Psi=-2^{n-3}\Sl_{\W p\in S_n}(-1)^{\W p}{W_I\W{U_J}\cdots
\W{U_K}\over z_{\W p}}~~~,~~~ z_p:=z_Iz_J\cdots
z_K~,~~~\label{eq:ExpandPf3} \eea
and
\bea \Pf~\Psi_m=(-1)^{{1\over 2}m(m+1)}\Sl_{\W p\in S_m}(-1)^{\W p}
{\W{U_I}\W{U_J}\cdots \W{U_K}\over z_{\W p}}~~~,~~~ z_p:=
z_Iz_J\cdots z_K~.~~~\label{eq:ExpandPf4} \eea
Here we sum over all possible partitions of $m$ elements into
subsets and for given partition, we sum over reflection independent
permutations for each subset. Remember that $\W U_I=U_I$ when $I$
only contains one or two elements.  For example, if $m=4$, the
cycles $\W p$ of $\Pf~\Psi_m$ are given by
\bea
\begin{array}{l}
\Bigl\{(1)(234)~,~(2)(134)~,~(3)(124)~,~(4)(123)~,~(1)(2)(34)~,~(1)(3)(24)~,~
(1)(4)(23)~,~(2)(3)(14)~,~~~\\
(2)(4)(13)~,~(3)(4)(12)~,~(12)(34)~,~(13)(24)~,~(14)(23)~,~(1234),
(1243)~,~(1324)~,~(1)(2)(3)(4) \Bigr\}
\end{array}~.~~~\label{eq:Cycles4pt} \eea
In the rest of this paper, we will always mention the $U$-cycles as
the $\W U$-cycles and use the rearranged expansions
\eqref{eq:ExpandPf3} and \eqref{eq:ExpandPf4}.
\begin{figure}
\centering
\includegraphics[width=3.7in]{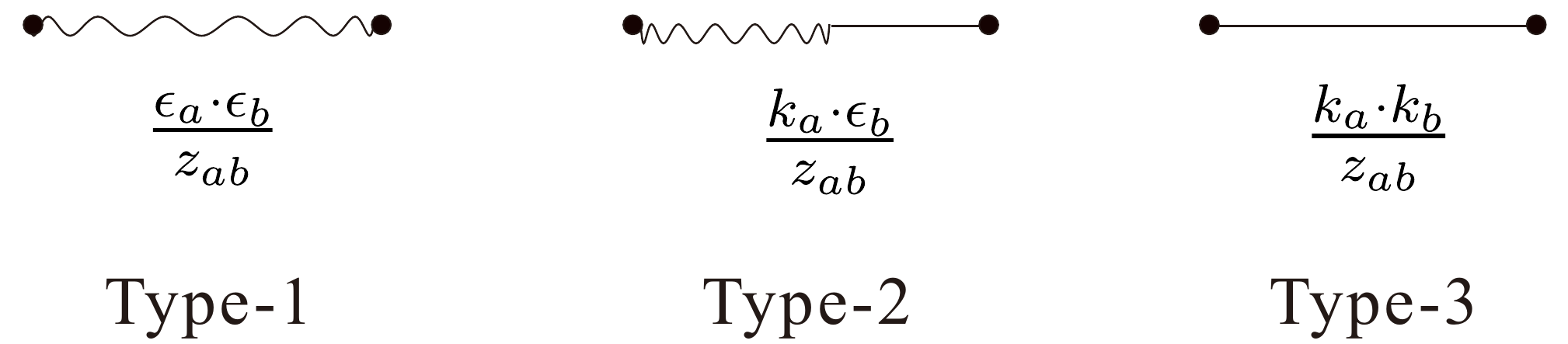}
\caption{Three types of lines.}\label{Fig:ThreeTypesOfLines}
\end{figure}
\begin{figure}
\centering
\includegraphics[width=2.5in]{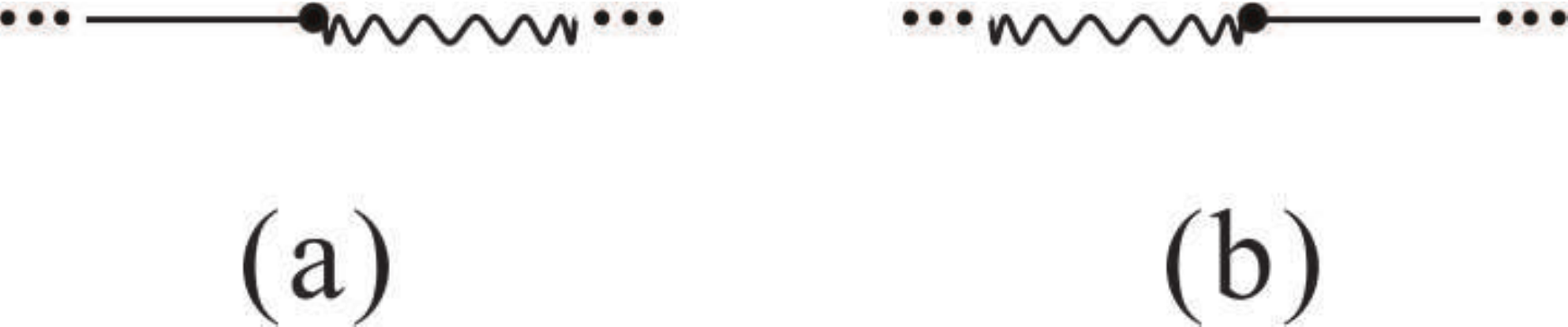}
\caption{If a node i) belongs to a $W$-cycle but is not an end of $W$-cycle or ii)
belongs to an $U$-cycle with more than one element, it gets contribution from the
 corresponding $W$- or $U$-cycle as shown by Figures (a) and (b). The two structures are related
 by flipping a sign because they are corresponding to the $p^{\mu}\epsilon^{\nu}$ and
 $-p^{\nu}\epsilon^{\mu}$ of $F^{\mu\nu}$.}\label{Fig:NodesInWorUCycle}
\end{figure}
%
\subsection{Diagrammatic rules}

Now let us establish the diagrammatic rules for writing Pfaffian or
reduced Pfaffian explicitly. To do this, we expand each $W$ and $\W
U$-cycle in terms of products of factors $(\epsilon\cdot\epsilon)$,
$(k\cdot k)$ and $(\epsilon\cdot k)$. A diagrammatic interpretation
for this expansion can be established as follows,
\begin{itemize}
\item We associate nodes with external particles. Two nodes $a$ and $b$
can be connected by (1) type-1 line if we have
$(\epsilon_a\cdot\epsilon_b)\over z_{ab}$, or (2) type-2 line if
we have $(\epsilon_a\cdot k_b)\over z_{ab}$, or (3) type-3 line
 if we have $(k_a\cdot k_b)\over z_{ab}$, as shown in Figure
 \ref{Fig:ThreeTypesOfNodes}. In this definition, the direction
of lines would matter and we will fix the convention of
direction later.

\item {\sl Contributions from $W$-cycle:} terms of a $W$-cycle always have two ends.
The two nodes play as the ends of $W$-cycle should be connected
with curved lines, i.e., type-1 line or the curved part of
type-2 line. This means if one end of such a line is node $a$,
we only have $(\epsilon_a\cdot \epsilon_i)$ or $(\epsilon_a\cdot
k_i)$ but do not have $(k_a\cdot\epsilon_i)$ and $(k_a\cdot
k_b)$. Other nodes on $W$-cycle between the two ends get
contributions which are shown by Figure
\ref{Fig:NodesInWorUCycle}. We should also have another type of
line connecting the two nodes $a$ and $b$, which represents
${1\over z_{ba}}$ (although in this paper, we will not deal with
$W$-cycle).

\item {\sl Contributions from $U$-cycles with more than one elements:}
an $U$-cycle with more than one element produces loop
structures. Each node belongs to an $U$-cycle also gets two
kinds of contributions from this cycle, as shown in Figure
\ref{Fig:NodesInWorUCycle}.b. An important point is that the two
lines connecting to the node must be one straight line and one
wavy line. In the definition of $\W U$ \eqref{WU-def}, we have
required that there are at least three elements. When there are
only two elements, we have instead
\bea {1\over 2} \Tr\big( (k_a \eps_a-\eps_a k_a)(k_b
\eps_b-\eps_b k_b)\big)=(\eps_a\cdot k_b)(\eps_b\cdot
k_a)-(\eps_a\cdot \eps_b)(k_b\cdot k_a)~.\eea
The disappearance of factor ${1\over 2}$ is the reason that we
can treat $U$-cycle with at least two elements uniformly.
Another thing is that, the $U$-cycle contains many terms with
relative $\pm $ sign. The diagrams with only type-2 lines will
have $(+)$ sign, and the sign of others shall be determined from
it. We will address the sign rule soon after.

\item {\sl Contributions from $U$-cycles with only one element:}
if a node $a$ belongs to a $U$-cycle with only one element
(i.e., $C_{aa}$), it could be connected with all other nodes via
$(\epsilon_a\cdot k_i)\over z_{ai}$. More precisely speaking,
using \eqref{Caa} one line connecting node $a$ and $i$ from
$C_{aa}$ should be $\left(z_{it}\over
z_{at}\right){(k_i\cdot\epsilon_a)\over z_{ia}}$, where $t$ is
the gauge choice. Thus this cycle contributes type-2 lines whose
curved part is connected to node $a$, multiplied by a factor
$\left(z_{it}\over z_{at}\right)$. This type of cycles can
contribute to either loop structure or tree structure.

\item {\sl Directions of lines:} for a loop diagram, we
read it clockwise.  For tree structures (which coming from
$C_{aa}$) connected to loop diagrams, we always read a (type-2)
line from the straight part ($k$) to the curved part
($\epsilon$).

\item {\sl Overall signs:} remember that each cycle is associated by a factor
$1$ when it contains odd number of elements and $(-1)$ when it
contains even number of elements. This is the overall sign.

\end{itemize}
With this diagrammatic interpretation, Pfaffian can be expanded as
tree structures rooted at loops. This diagrammatic rule can be
regarded as a generalization of spanning tree expression for MHV
gravity \cite{Feng:2012sy} and EYM amplitudes \cite{Du:2016wkt}.

\subsection{Examples}

%
\begin{figure}
\centering
\includegraphics[width=5.5in]{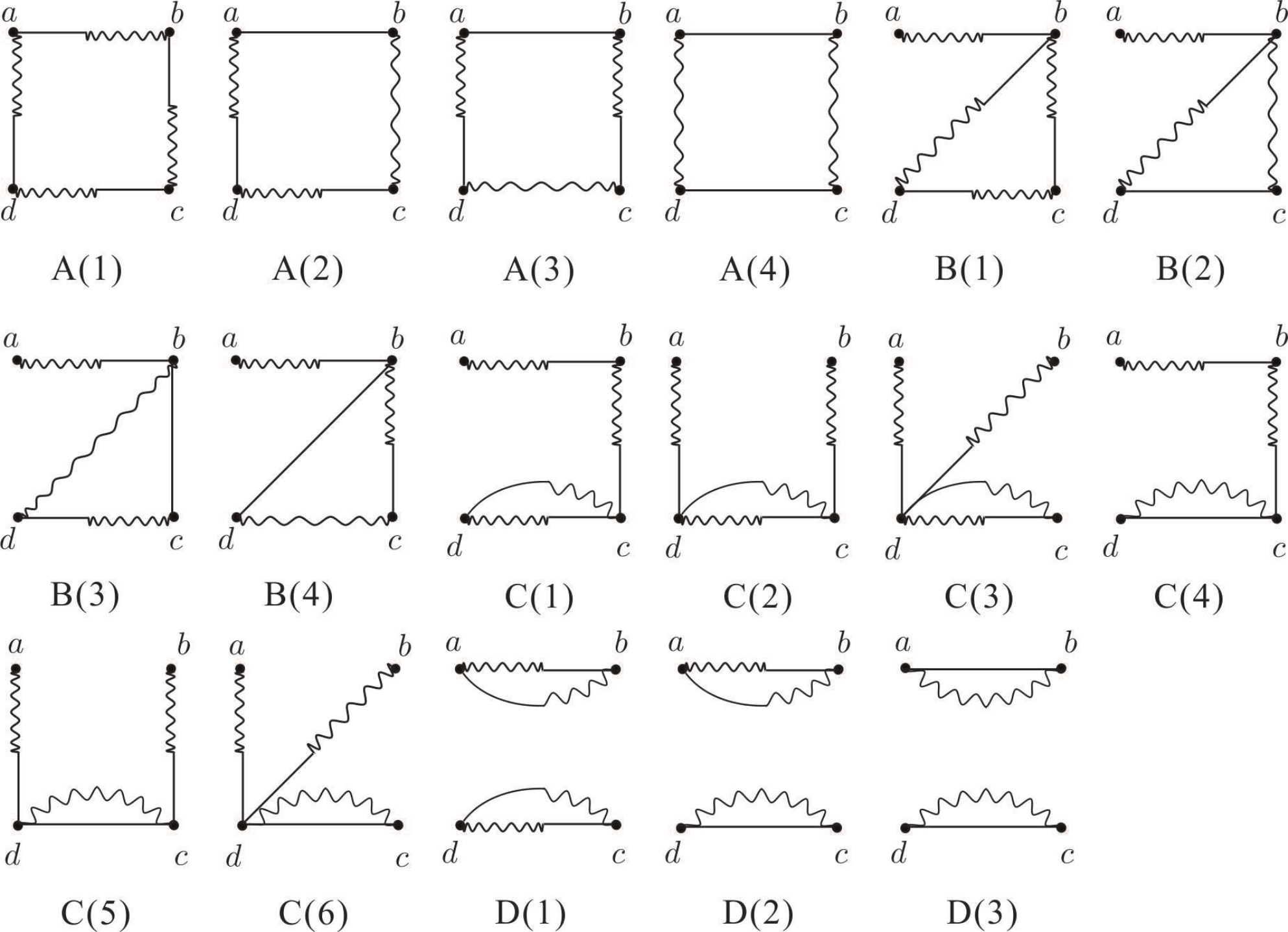}
\caption{All possible structures in the four-element example.}\label{Fig:FOURptDiagrams}
\end{figure}
Now let us take the expansion of Pfaffian $\text{Pf}\,(\Psi_4)$ as
an example to illustrate. There are five types of cycles: $(abcd)$,
$(a)(bcd)$, $(ab)(cd)$,  $(a)(b)(cd)$ and $(a)(b)(c)(d)$, where
$a,b,c,d$ can label as permutations of $1,2,3,4$. All reflection
independent cycles are already given by \eqref{eq:Cycles4pt}. The
$(abcd)$ contains only $U$-cycle with more than one element, while
the $(a)(b)(c)(d)$ only gets contribution from $C_{aa}$'s. We
consider these cycles one by one.

For the $U$-cycle $(1234)$, we have four possible structures, as
shown by the diagrams Figure \ref{Fig:FOURptDiagrams}. A(1), A(2),
A(3) and A(4). We consider each diagram as a function of
$\epsilon_a, k_a$, $\epsilon_b, k_b$, $\epsilon_c, k_c$ and
$\epsilon_d, k_d$, and denote e.g, A(1) by
\bea A_1(abcd):=(k_a\cdot \eps_b)(k_b\cdot \eps_c)(k_c\cdot
\eps_d)(k_d\cdot \eps_a)~.~~~\label{A1-abcd}\eea
With this notation,  $\W U(1234)$ is given by
\bea
\W U(1234)&=&(-1)\Bigl[A_{1}(1234)-A_{2}(1234)-A_{2}(4123)-A_{2}(3412)-A_{2}(2341)\nonumber\\
&&+A_{1}(1432)-A_{2}(1432)-A_{2}(2143)-A_{2}(3214)-A_{2}(4321)\nn
&&+A_{3}(1234)+A_{3}(4123)+A_{3}(3412)+A_{3}(2341)+A_{4}A(1234)+A_{4}(4123)\Bigr]~.~~~\label{U1234-exp}\eea
Let us pause a little bit to explain \eqref{U1234-exp}. With
$\Tr(FFFF)$, after expanding  we will get 16 terms as in
\eqref{U1234-exp}. However, some terms share the same pattern and in
current case, there are four patterns. Now we present a trick to
find these patterns for a loop diagram:
\begin{itemize}

\item First let us assign a number to three types of lines: $0$ for the
type-1, $1$ for the type-2 and $2$ for the type-3. In fact, this
number is the mass dimension of these lines. With this
assignment, we can write down the cyclic ordered lists for $A_i$
as
\bea A_1\to (1,1,1,1)~~~,~~~A_2\to (2,0,1,1)~~~,~~~A_3\to
(2,1,0,1)~~~,~~~A_4\to (2,0,2,0)~.~~~\eea

\item Now we can see the construction of patterns for $\W U(1234)$.
First we split $4$ into four number $n_i$ to construct the
ordered list $(n_1,n_2, n_3, n_4)$, such that: (1) $n_i\in
\{0,1,2\}$, (2) $\sum_{i=1}^4 n_i=4$, (3) If $n_i=2$, then both
$n_{i-1}, n_{i+1}$ can not be $2$. Similarly If $n_i=0$, then
both $n_{i-1}, n_{i+1}$ can not be $0$. After getting the
ordered list, we compare them. If  two ordered list $(n_1,n_2,
n_3, n_4)$ and $(\W n_1,\W n_2, \W n_3, \W n_4)$ are the same
either by cyclic rotation or by order-reversing, we will say
they have defined the same pattern.

With above rule, it is easy to see that, (1) When there are two
$n_i$ taking value $2$, the only allowed list is
$A_4:(2,0,2,0)$, (2) When there is only one $n_i$ taking value
$2$, for example, $n_2=2$, there are four possibilities
$(0,2,0,\ast)$, $(1,2,1,\ast)$, $(0,2,1,\ast)$ and
$(1,2,0,\ast)$. However, the sum to be $4$ picks only the latter
three $(1,2,1,0)$, $(0,2,1,1)$ and $(1,2,0,1)$. Since the last
two are related by order-reversing, we get the patterns
$A_3:(1,2,1,0)$ and $A_2:(1,2,0,1)$, (3) When there is no
$n_i=2$, the only possibility is $A_1:(1,1,1,1)$.

\item In fact, we can get all patterns and their relative sign
starting from the fundamental pattern $A_1:
+(1,1,1,1)$\footnote{The fundamental pattern is the one with all
$n_i=1$ and its sign is always $+1$. One must be careful for the
rule of type-2 line. } by the so called {\sl flipping action}.
The flipping action is defined as taking two nearby
$(n_i,n_{i+1})$ and changing it to $(n_i-1, n_{i+1}+1)$ or
$(n_i+1, n_{i-1}+1)$. It is wroth to notice that the allowed
flipping action must satisfy that (1) obtain new $n_i\in
\{0,1,2\}$, (2) no two $2$ or two $0$ are nearby. If a pattern
is obtained from fundamental pattern by odd number of flipping
actions, its sign is negative, while if a pattern is obtained
from fundamental pattern by even number of flipping actions, its
sign is positive.

Using above rule, it is easy to see that the sign for $A_2$ is
$(-)$ and for $A_3, A_4$, $(+)$.

\end{itemize}

Having done the $(1234)$, we move to the $(1)(234)$ case and the
result is
\bea \W U(1)\W U(234)&=&{z_{1t}\over
z_{2t}}\Bigl[B(1)(1234)+B(1)(1243) -B(2)(1234)-B(2)(1243)\nn
&&-B(3)(1234)-B(3)(1243)-B(4)(1234)-B(4)(1243)\Bigr]+\cyclic{2,3,4}~.~~~
\eea
Again let us give some explanations. Unlike the case $(1234)$,
because of the cycle $(1)$, the node of $2,3,4$ connecting to node
$1$ is special, so cyclic symmetry has lost although the order
reversing symmetry is still kept. Using the algorithm, we split $3$
into three ordered positions to get $B_1=(1,1,1)$, $B_4=(2,1,0),
B_3=(0,2,1),B_2=(1,0,2)$ (another three $(2,0,1),(1,2,0),(0,1,2)$
are order reversing comparing to previous three, so we do not
count). Furthermore, $B_2, B_3, B_4$ are obtained from $B_1$ by one
flipping action, so their relative sign is $(-)$.

For the third case $(12)(34)$, the result is
\bea &&\W U(12)\W
U(34)=D(1)(1234)-D(2)(3412)-D(2)(1234)+D(3)(1234)~.~~~ \eea
Since each $(ab)$ gives $+(1,1), -(2,0)$ patterns, when multiplying
together, we get $D_1=+(1,1)(1,1), D_2=-(1,1)(2,0),D_3=+(2,0)(2,0)$
three patterns as shown in Figure \ref{Fig:FOURptDiagrams}.

The fourth case $(1)(2)(3)(4)$ is a little bit different. Unlike the
loop diagram (i.e., cycle with at least two elements) with three
types of lines, here we can have only type-2 line for single cycle
$C_{aa}$. Thus the problem  is reduced to find the $2$-regular
graph, i.e., each node has two and only two lines connecting to it.
Thus there are only two patterns: $D_1$ and $A_1$. One complication
for the single cycle is that there is an extra factor ${z_{it}\over
z_{at}}$ attaching to the type-2 line.

For the last case $(a)(b)(cd)$, the situation is the most
complicated. The $(cd)$ cycle gives $(1,1)$ and $(2,0)$ two
patterns, but depending on how single cycles are attached, we can
have (1) for $a,b$ attached to each other,  it reduces to $D_1, D_2,
D_3$, (2) for $a$ attached to $b$, but $b$ attached to, for example,
$c$, it gives $C_1, C_4$, (3) for $a,b$ attached to same point, for
example, $d$, it gives $C_3, C_6$, (4) for $a,b$ attached to
different points, it gives $C_2, C_5$.

\section{The cancelation of double poles in Yang-Mills theory and gravity}
\label{secCancelation}

In this section,  we investigate the cancelation of higher-order
poles in Yang-Mills and gravity theories. The building blocks of
these two theories are PT-factor and the reduced Pfaffian. Since for
PT-factor, we always have $\chi_L(A)\leq 0$, and the trouble comes
from the reduced Pfaffian, where $\chi_R(A)=1$ do appear. For
instance, if we consider higher-order pole with three elements
$\{a,b,c\}$\footnote{We will not consider the case where elements of
the subset $A$ have been chosen as gauge for reduced Pfaffian. We
will discuss this situation later. }, we only need to consider the
terms with cycles $(abc)$, $(a)(bc)$, $(b)(ac)$, $(c)(ab)$ and
$(a)(b)(c)$. When summing together, it is easy to see that these
terms having the form $\cdots {\rm Pf}(\Psi_{abc})$. This pattern is
general, thus for possible double pole $s_A$, our focus will be
${\rm Pf}(\Psi_{A})$. We will show by some examples that after using
various on-shell and off-shell identities, ${\rm Pf}(\Psi_{A})$
could effectively have $\chi(A)=0$, by either terms with explicit
$\chi=1$ having numerator factor $s_A$ or when summing some terms
together, the $\chi=1$ is reduced to $\chi=0$.

\subsection{The cancelation of higher-order poles with two elements}
%
\begin{figure}
\centering
\includegraphics[width=3in]{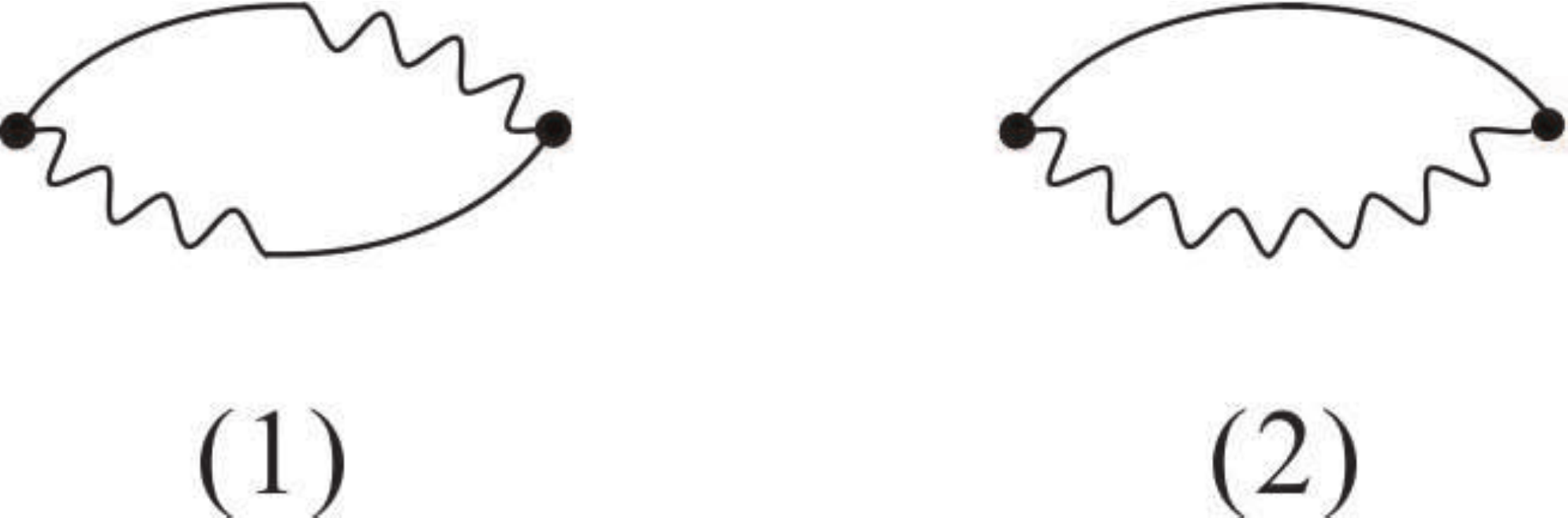}
\caption{Diagrams contributing to double pole in $s_{12}$ channel.}\label{Fig:2ptDoublePole}
\end{figure}
This is the first non-trivial case, and we will study it from
different approaches to clarify some conceptual points.

Let us start with explicit evaluation of ${\rm Pf}(\Psi_{12})$.
There are two cycles $(12)$ and $(1)(2)$. For $(12)$ cycle, the
contribution is
\bea { {1\over 2}{\rm Tr}( (k_1\eps_1-\eps_1 k_1)(k_2\eps_2-\eps_2
k_2)) \over \spaa{12}}={ (k_1\cdot \eps_2) (k_2\cdot
\eps_1)-(k_1\cdot k_2) (\eps_1\cdot \eps_2)\over
\spaa{12}}~,~~~\label{Pole12-(12)}\eea
where for simplicity the notation \eref{notation} has been applied.
For the cycle $(1)(2)$, when using \eqref{Caa}, we take the same
gauge choice $t=n$, and get
\bea \left\{(\eps_1\cdot k_2){ z_{2t}\over z_{21}z_{1t}}+\sum_{j=3}^{n-1}
(\eps_1\cdot k_j){ z_{jt}\over z_{j1}z_{1t}}\right\} \left\{(\eps_2\cdot
k_1){ z_{1t}\over z_{12}z_{2t}}+\sum_{q=3}^{n-1} (\eps_2\cdot k_q){
z_{qt}\over z_{q2}z_{2t}}\right\}~.~~~\eea
It is easy to see that the other terms will have $\chi(\{1,2\})\leq
0$, except the following part
\bea \left\{(\eps_1\cdot k_2){ z_{2t}\over z_{21}z_{1t}}\right\}
\left\{(\eps_2\cdot k_1){ z_{1t}\over z_{12}z_{2t}}\right\}= {(k_1\cdot
\eps_2) (k_2\cdot \eps_1)\over
\spaa{12}}~.~~~\label{Pole12-(1)(2)-1} \eea
Now we need to combine these two terms with the proper sign and get
\bea {(k_1\cdot \eps_2) (k_2\cdot \eps_1)\over \spaa{12}}-{
(k_1\cdot \eps_2) (k_2\cdot \eps_1)-(k_1\cdot k_2) (\eps_1\cdot
\eps_2)\over \spaa{12}}={(k_1\cdot k_2) (\eps_1\cdot \eps_2)\over
\spaa{12}} ~.~~~\label{Pole12-(1)(2)-2} \eea
We see immediately that, although the denominator $\spaa{12}$ gives
$\chi(\{1,2\})=1$, the explicit numerator factor $(k_1\cdot
k_2)={1\over 2} s_{12}$ will reduce double pole to single pole.

Above calculation is correct but a little too rough.  We need to
show that the result should not depend on the gauge choice for the
single cycle.  Now let us present a systematic discussion on this
issue,
\begin{itemize}

\item Firstly, from the expansion \eqref{Caa} we see that there are
two choices for the gauge $t$. In the first choice, we choose
$t\in A$. In this case, no matter which $j$ is, the linking
number is always $+1$, so we need to sum over all $j$. In the
second choice, we choose $t\not \in A$, thus only when $j\in A$,
we get the linking number one which contributes to double pole.
This tells us that to simplify the calculation, we should take
$t\not \in A$.

\item In our previous calculation, although we have taken $t\not\in
A$, we have made the special choice to set the same $t$ for both
$C_{11}, C_{22}$. In general we could take two different gauge
choices, so we are left with(again, with such gauge choice, only
those $j\in A$ are needed to be sum over)
\bea & & \left\{(\eps_1\cdot k_2){ z_{2t}\over z_{21}z_{1t}}\right\}
\left\{(\eps_2\cdot k_1){ z_{1\W t}\over z_{12}z_{2\W t}}\right\}=
{(k_1\cdot \eps_2) (k_2\cdot \eps_1)\over \spaa{12}} {z_{2t}z_{1\W
t}\over z_{1 t}z_{2\W t}}\nn
& = &{(k_1\cdot \eps_2) (k_2\cdot \eps_1)\over \spaa{12}} {z_{2\W
t}z_{1 t}+z_{21}z_{t \W t}\over z_{1 t}z_{2\W t}} \to  {(k_1\cdot \eps_2)
(k_2\cdot \eps_1)\over \spaa{12}}~.~~~\label{Pole12-(1)(2)-3}
\eea
Among the two terms at the second line, since the numerator
$z_{21}$ in the second term has decreased the linking number by
one, we are left with only the first term, which is the same
result as \eref{Pole12-(1)(2)-1}.

\item Now we consider the gauge choice $t\in A$,
for example $t=1$ for $C_{22}$. Then we will have
\bea & & \left\{ (\eps_1\cdot k_2) { z_{2t}\over z_{21}z_{1t}}\right\}
\left\{ \sum_{j=3}^n (\eps_2\cdot k_j){ z_{j1}\over
z_{j2}z_{21}}\right\}\nn
& = & \sum_{j=3}^n {(\eps_1\cdot k_2)  (\eps_2\cdot k_j)\over
-\spaa{12}} {z_{j1}z_{2t}\over z_{j2}z_{1t}}=\sum_{j=3}^n {(\eps_1\cdot
k_2) (\eps_2\cdot k_j)\over -\spaa{12}} {z_{j2}z_{1t}+z_{jt}z_{21}\over
z_{j2}z_{1t}}~.~~~\eea
Again, after dropping the second term, we are left with
\bea \sum_{j=3}^n {(\eps_1\cdot k_2) (\eps_2\cdot k_j)\over
-\spaa{12}}=  {(\eps_1\cdot k_2) (\eps_2\cdot k_1)\over
\spaa{12}}~,~~~\eea
which is the same result as \eref{Pole12-(1)(2)-1}.

\end{itemize}
By above detailed discussions, we see that after properly using the
various (such as Schouten) identities, momentum conservation and
on-shell conditions, we do get the same answer for arbitrary gauge
choices. With this clarification, in the latter computations we will
take proper gauge choice without worrying the independence with the
gauge choice.

Now we will use our diagrammatic rules to re-do above calculation.
The purpose of presenting both calculations is to get familiar with
our new technique and find the general pattern for later examples.
The potential contribution of double poles with two elements  $1$
and $2$ comes from the cycles $(12)$ and $(1)(2)$. There are two
kinds of diagrams (see Figure \ref{Fig:2ptDoublePole})\footnote{As
we have explained above, for single cycle $C_{aa}$ to contribute to
$\chi(A)=1$, we must have $j\in A$, which can be seen clearly if we
take gauge $t\not\in A$. Thus for cycle $(1)(2)$, we need to
consider only the case when node $1$ is connected to node $2$.}. The
first diagram gets contribution from both $(12)$ cycle and $(1)(2)$
cycle. Particularly, it reads (noticing that each two-element cycle
contains a $(-1)$ and one element cycle contains $1$)
\bea -{(k_1\cdot\epsilon_2)(k_2\cdot\epsilon_1)\over
\spaa{12}}+{z_{1t}\over z_{2t}}{(k_1\cdot\epsilon_2)\over
z_{12}}{z_{2t}\over z_{1t}}{(k_2\cdot\epsilon_1)\over
z_{21}}=0~,~~~\eea
where we choose the same gauge $z_t$ for $C_{11}$ and $C_{22}$. The
second diagram evaluates to
\bea {(k_1\cdot k_2)(\epsilon_1\cdot\epsilon_2)\over \spaa{12}}~.~~~
\eea
Thus we have simply reproduced \eref{Pole12-(1)(2)-1}.

\subsection{The cancelation of higher-order pole with three elements}
%
\begin{figure}
\centering
\includegraphics[width=2.7in]{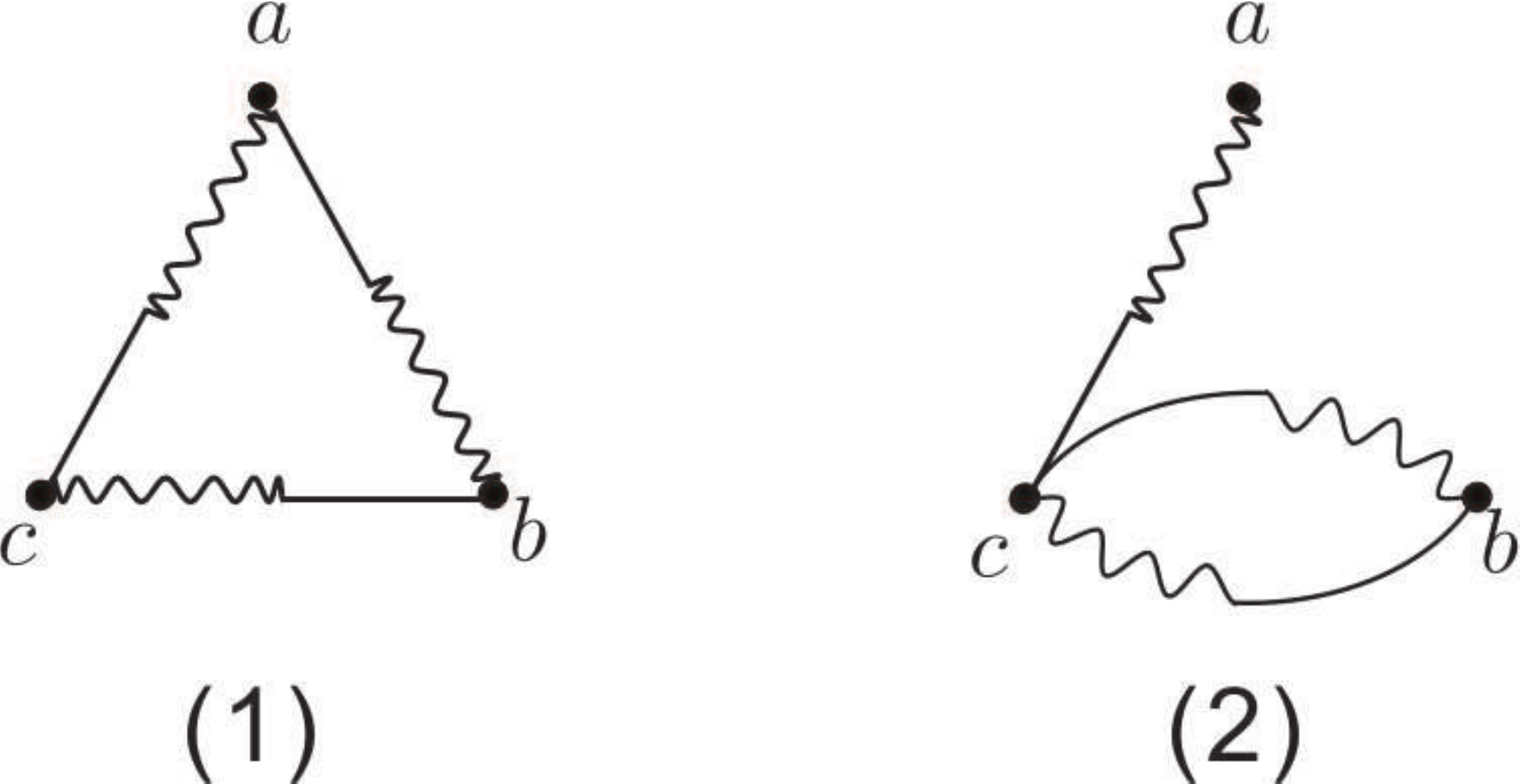}
\caption{Diagrams contributing to double pole, which only contain type-2 lines. }\label{Fig:3ptDoublePole}
\end{figure}
\begin{figure}
\centering
\includegraphics[width=2.7in]{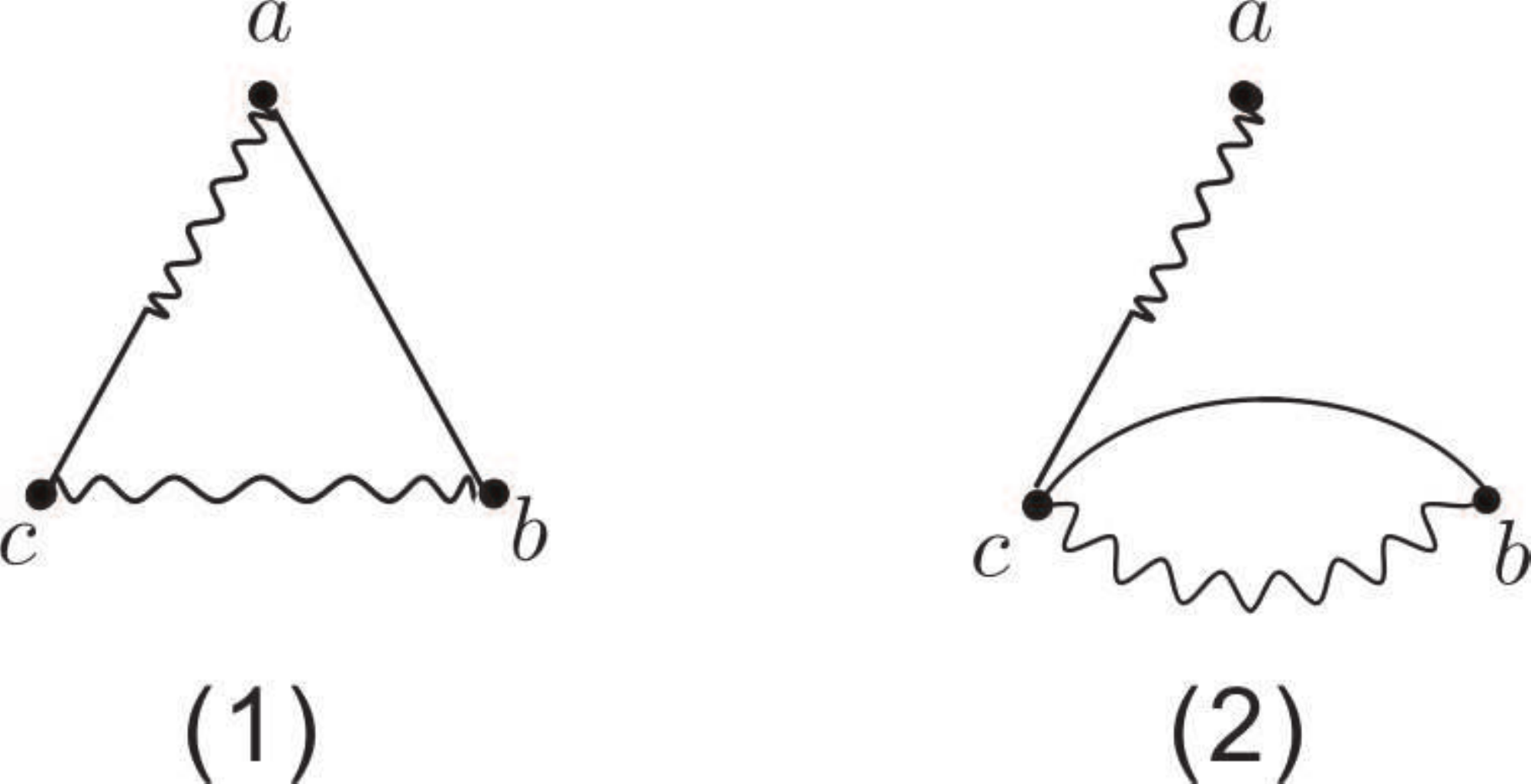}
\caption{Diagrams contributing to double pole, which contain type-1 and type-3 lines.}\label{Fig:3ptDoublePole2}
\end{figure}
Now let us consider the cancelation of double poles with three
elements using the diagrammatic rules developed in this paper. There
are cycles $(123)$, $(1)(23)$, $(12)(3)$, $(13)(2)$ and $(1)(2)(3)$
contributes. We collect their contributions according to the pattern
of kinematic factors,
\begin{itemize}
\item Diagrams containing at least one type-2 loop (loops constructed by only type-2 lines)
are shown by Figure \ref{Fig:3ptDoublePole}. This is the
complicated case since both cases, i.e., $U$-cycle with at least
two elements and single cycle merging, will contribute. Thus
Figure \ref{Fig:3ptDoublePole}.1 gets contribution from $(123)$
and $(1)(2)(3)$ cycles and can be evaluated to
\bea
    -{(k_a\cdot\epsilon_b)(k_b\cdot\epsilon_c)(k_c\cdot\epsilon_a)\over
    \spaa{abc}}+{ z_{at}\over z_{bt}}{(k_a\cdot\epsilon_b)\over
     z_{ab}}{ z_{bt}\over z_{ct}}{(k_b\cdot\epsilon_c)\over
     z_{bc}}{ z_{ct}\over z_{at}}{(k_c\cdot\epsilon_a)\over
     z_{ca}}=0~.~~~\eea

Similarly, Figure \ref{Fig:3ptDoublePole}.2 gets contribution
from $(a)(bc)$ and $(a)(b)(c)$ cycles and can be evaluated to
\bea {(\epsilon_a\cdot k_c)\over  z_{ac}}\left[{
-{(k_a\cdot\epsilon_b)(k_b\cdot\epsilon_a) \over \spaa{bc}}}+{
z_{at}\over  z_{bt}}{(k_a\cdot\epsilon_b)\over z_{ab}}{
 z_{bt}\over  z_{at}}{(k_b\cdot\epsilon_a)\over
 z_{ba}}\right]=0~.~~~\eea
Thus all diagrams containing a type-2 loop are canceled.

\item Diagrams do not contain any type-2 loop. In this case,
 all loop structures  should also contain type-1 and type-3
lines. For the case with three elements, we have two kinds of
typical diagrams, as shown in Figure \ref{Fig:3ptDoublePole2}.
The first one comes from the cycle $(abc)$ while the second
diagram comes from $(a)(bc)$. Thus the two diagrams in Figure
\ref{Fig:3ptDoublePole2} gives
    \bea & & -{\epsilon_a\cdot k_c\over
     z_{ac}}{\epsilon_c\cdot\epsilon_b\over
     z_{cb}}\left[{k_b\cdot k_a\over z_{ba}}+{ z_{ct}\over
     z_{at}}{k_b\cdot
    k_c\over  z_{bc}}\right]=-{\epsilon_a\cdot k_c\over
     z_{ac}}{\epsilon_c\cdot\epsilon_b\over
     z_{cb}}\left[{k_b\cdot k_a\over z_{ba}}\left(1+{k_b\cdot
     k_c\over k_b\cdot
    k_a}{ z_{ba}\over  z_{bc}}{ z_{ct}\over
     z_{at}}\right)\right]\nn
    &=& {\epsilon_a\cdot k_c\over
     z_{ac}}{\epsilon_c\cdot\epsilon_b\over
     z_{cb}}\left[{k_b\cdot k_a\over  z_{ba}}\Sl_{i\neq
    a,b,c,t} {k_b\cdot k_i\over k_b\cdot k_a}{ z_{ba}\over
     z_{bi}}
    { z_{it}\over z_{at}}\right]={\epsilon_a\cdot k_c\over
    z_{ac}}{\epsilon_c\cdot\epsilon_b\over
     z_{cb}}\left[\Sl_{i\neq a,b,c,t} {k_b\cdot k_i z_{it}\over
    z_{bi}  z_{at}}\right]~,~~~ \eea
where the cross-ratio identity \eqref{crossratio}  has been used
with the subset $A=\{a,b\}$ and gauge choice $(a,t)$. The part
inside the bracket has the following features, (1) cancelation
of $(k_a\cdot k_b)$ between numerator and denominator, (2)
cancelation of $z_{ba}$ between numerator and denominator. In
the final form, the RHS of above expression is still weight-2
graph for all nodes, but  the linking number contribution is
effectively reduced by one, i.e., $\mathbb{L}(\{a,b,c\})=3-1=2$
by $z_{ac}z_{cb}$.

\end{itemize}

Before finishing this subsection, let us compare this example with
the one in the previous subsection. These two examples have shown
two different patterns of removing double poles. In the first
example, it is the explicit numerator factor $s_{ab}$ that removes
the double pole but the linking number is not changed. In the second
example, after using the cross-ratio identity, linking number is
effectively decreased by one but there is no $s_{abc}$ factor in the
numerator.

\subsection{The cancelation of higher-order poles with four elements}

It is natural to generalize the discussions above to more
complicated cases, which can be summarized as, (1) all diagrams
containing at least one type-2 loop should be canceled, (2) the
other diagrams (containing type-1 and type-3 lines) are grouped
together if they give the same diagram when all type-3 lines in them
are removed (e.g., the two diagrams in Figure
\ref{Fig:3ptDoublePole2}). The cancelation of double poles in these
diagrams are results of cross-ratio identity. Now let us take the
cancelation of double poles with four elements as a more general
example to see these two kinds of cancelations.

\subsubsection*{The cancelation between diagrams containing type-2 loops}

A pure type-2 loop can come from either $U$-cycle with more than one
elements or a product of $U$-cycles each contains one element. In
the four-element case, the diagrams containing type-2 loops are
given by (A1), (B1), (C1), (C2), (C3), (D1) and (D2) in Figure
\ref{Fig:FOURptDiagrams}. We take the (A1) diagram in Figure
\ref{Fig:FOURptDiagrams} as an example. The Figure
\ref{Fig:FOURptDiagrams}.A1 receives  the following contribution
from $U$-cycle $(abcd)$ with four elements,
\bea -{1\over \spaa{abcd}}(k_a\cdot \epsilon_b)(k_b\cdot
\epsilon_c)(k_c\cdot \epsilon_d)(k_d\cdot \epsilon_a)~,~~~ \eea
and a contribution from $U$-cycles $(a)(b)(c)(d)$, each containing
one element, i.e.,
\bea {1\over \spaa{abcd}} { z_{at}\over z_{bt}}(k_a\cdot
\epsilon_b){ z_{bt} \over z_{ct}}(k_b\cdot \epsilon_c){ z_{ct}\over
z_{dt}} (k_c\cdot \epsilon_d){ z_{dt}\over z_{at}}(k_d\cdot
\epsilon_a) ={1\over \spaa{abcd}}(k_a\cdot \epsilon_b)(k_b\cdot
\epsilon_c)(k_c\cdot \epsilon_d) (k_d\cdot \epsilon_a)~,~~~\eea
where all one-element cycles taking the same gauge choice $t\neq
a,b,c,d$. Apparently, these two contributions are canceled with each
other.

This cancelation is easily generalized to cases containing at least
one type-2 loop. If we consider a diagram containing a type-2 loop
with nodes $a_1$, $a_2$, $\ldots, a_m$ on it, the $U$-cycle with $m$
elements $(a_1a_2\cdots a_m)$ and the product of $m$ one-element
$U$-cycles $(a_i),i=1,\ldots,m$) contribute to this loop. The
$U$-cycle $(a_1a_2\cdots a_m)$ contribution is written as
\bea (-1)^{m+1}{1\over \spaa{a_1a_2\cdots
a_m}}(-1)^m\left(k_{a_1}\cdot\epsilon_{a_2}
\right)\left(k_{a_2}\cdot\epsilon_{a_3}\right)\cdots
\left(k_{a_m}\cdot\epsilon_{a_1}\right)~,~~~ \eea
where the first pre-factor $(-1)^{m+1}$ comes from the pre-factor in
front  of $U$-cycle, i.e., $-1$ for even number of elements, while
$1$ for odd number of elements. The second factor $(-1)^{m}$ comes
from the contribution from the expansion of $U$-cycle, only the term
with a minus in each $F_i^{\mu\nu}$ factor contributes. The product
of one-element $U$-cycles $C_{a_1a_1}C_{a_2a_2}\cdots C_{a_ma_m}$
contributes a
\bea &&{1\over \spaa{a_1a_2\cdots a_m}} { z_{a_1t}\over
z_{a_2t}}(k_{a_1}\cdot \epsilon_{a_2}){ z_{a_2t}\over
z_{a_3t}}(k_{a_2}\cdot \epsilon_{a_3}) \cdots { z_{a_mt}\over
z_{a_1t}}(k_{a_m}\cdot \epsilon_{a_1})\nn
&=&{1\over \spaa{a_1a_2\cdots
a_m}}\left(k_{a_1}\cdot\epsilon_{a_2}\right)
\left(k_{a_2}\cdot\epsilon_{a_3}\right)\cdots
\left(k_{a_m}\cdot\epsilon_{a_1}\right)~,~~~\eea
where we have chosen the  $t$ ($t\neq a_1,\cdots, a_m$) for all
$C_{ii}$ to be the same. This expression is precisely canceled with
the corresponding contribution from $m$-element $U$-cycle. After
such cancelations, the diagrams (A1), (B1), (C1), (C2), (C3), (D1)
and (D2) in Figure \ref{Fig:FOURptDiagrams} are all canceled. Thus
only those diagrams which do not contain any type-2 loop survive.

\subsubsection*{The cancelation of double poles in diagrams which do not contain any type-2 loop}

Now let us turn to the diagrams with no type-2 loop. As shown in the
case of  three-element poles, we should group together those
diagrams which are the same after removing all type-3 lines. The
cancelation of double poles can be found by applying cross-ratio
identity. In the four-element case, we have the following types of
cancelations,

(1) The first type of cancelation happens between diagrams (A2),
(B4) and (C4) with respect to the cycles $(abcd)$, $(a)(dbc)$ and
$(a)(d)(bc)$. The potential contributions to four-element
higher-order poles are collected as
\bea \begin{array}{c}
\includegraphics[width=0.9\textwidth]{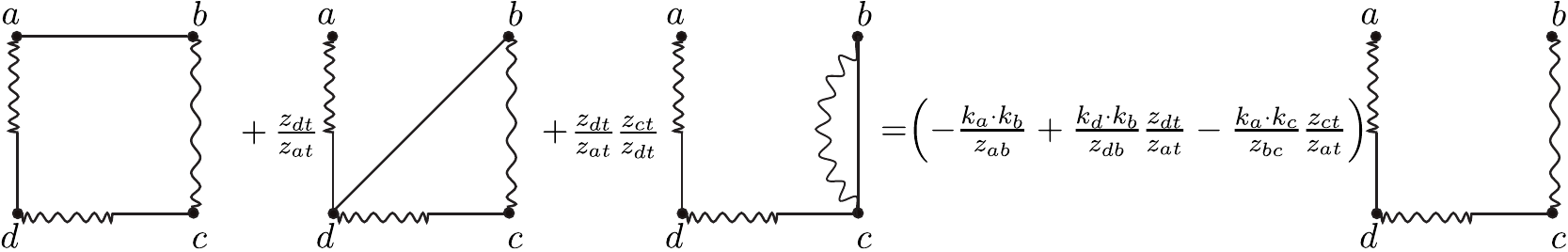}\end{array}~.~~~\label{eq:4ptDoublePole1}
\eea
The RHS of above equation reads
\bea &&\left(-\frac{k_a\cdot k_b}{ z_{ab}}+\frac{k_d\cdot k_b}{
z_{db}}\frac{ z_{dt}}{ z_{at}}-\frac{k_b\cdot k_c}{ z_{bc}}\frac{
z_{ct}}{ z_{at}}\right){\epsilon_b\cdot\epsilon_c\over
 z_{bc}}{k_c\cdot\epsilon_{d}\over  z_{cd}}{k_d\cdot
\epsilon_a\over  z_{da}}\nn
&=&-\frac{k_a\cdot k_b}{ z_{ab}}\left(1+{s_{bd} \over s_{ba}}\frac{
z_{ba}}{ z_{bd}}\frac{ z_{dt}}{ z_{at}}+\frac{s_{bc}}{s_{ba}}\frac{
z_{ba}}{ z_{bc}}\frac{ z_{ct}}{
z_{at}}\right){\epsilon_b\cdot\epsilon_c\over
 z_{bc}}{k_c\cdot\epsilon_{d}\over  z_{cd}}{k_d\cdot
\epsilon_a\over  z_{da}}~.~~~\eea
Applying cross-ratio identity \eqref{crossratio}, this contribution
becomes
\bea \frac{k_a\cdot k_b}{ z_{ab}}\left(\Sl_{e\neq a,b,c,d,t}{s_{be}
\over s_{ba}}\frac{ z_{ba}}{ z_{be}}\frac{ z_{et}}{ z_{at}}\right)
{\epsilon_a\cdot k_d\over  z_{ad}}{\epsilon_{d}\cdot k_c\over
 z_{dc}} {\epsilon_c\cdot\epsilon_b\over  z_{cb}}=
-\left(\Sl_{e\neq a,b,c,d,t}{k_b\cdot k_e  z_{et}\over
 z_{be} z_{at}}\right) {\epsilon_a\cdot k_d\over
 z_{ad}}{\epsilon_{d}\cdot k_c\over  z_{dc}}
{\epsilon_c\cdot\epsilon_b\over  z_{cb}}~,~~~\eea
in which, while keeping the weight-2 conditions for every nodes, the
linking number for this part is decreased to $m-1=4-1=3$.

(2) The second type of cancelation happens between diagrams (A3),
(B3) and (C5). Particularly, we have
\bea
\begin{array}{c}
\includegraphics[width=0.7\textwidth]{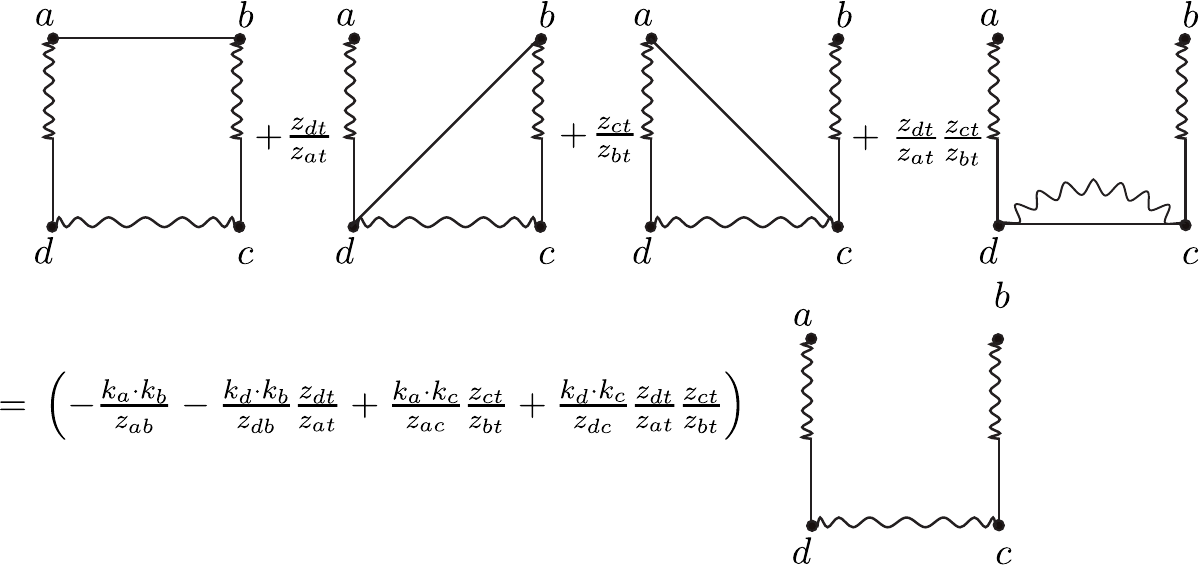}\end{array}~,~~~\label{eq:4ptDoublePole2}
\eea
whose RHS is explicitly written as
\bea \left[-\frac{k_a\cdot k_b}{ z_{ab}}\left(1+{s_{bd}\over
s_{ba}}{ z_{ba}\over z_{bd}}\frac{ z_{dt}}{ z_{at}}\right)+{k_a\cdot
k_c\over z_{ac}}{ z_{ct}\over z_{bt}}\left(1+{s_{cd}\over s_{ca}}{
z_{ca}\over
 z_{cd}}\frac{ z_{dt}}{ z_{at}}\right)\right]{\epsilon_a\cdot
k_d\over  z_{ad}}{\epsilon_d\cdot\epsilon_c\over
 z_{dc}}{k_c\cdot\epsilon_b\over
 z_{cb}}~.~~~\label{eq:4ptDoublePole2a} \eea
According to the cross-ratio identity \eqref{crossratio}, we have
\bea 1+{s_{bd}\over s_{ba}}{ z_{ba}\over z_{bd}}\frac{ z_{dt}}{
z_{at}}&=&-{s_{bc}\over s_{ba}}{ z_{ba}\over z_{bc}}\frac{ z_{ct}}{
z_{at}}-\Sl_{e\neq a,b,c,d,t}{s_{be}\over s_{ba}}{ z_{ba}\over
z_{be}}\frac{ z_{et}}{ z_{at}}~,~~~\\
1+{s_{cd}\over s_{ca}}{ z_{ca}\over
 z_{cd}}\frac{ z_{dt}}{ z_{at}}&=&-{s_{cb}\over
s_{ca}}{ z_{ca}\over
 z_{cb}}\frac{ z_{bt}}{ z_{at}}-\Sl_{e\neq
a,b,c,d,t}{s_{ce}\over s_{ca}}{ z_{ca}\over
 z_{ce}}\frac{ z_{et}}{ z_{at}}~.~~~\eea
Plugging these identities into \eqref{eq:4ptDoublePole2a}, we
immediately arrive at
\bea -{1\over 2}\left[\Sl_{e\neq a,b,c,d,t}{s_{be}\over
z_{be}}\frac{ z_{et}}{ z_{at}}-\Sl_{e\neq a,b,c,d,t}{s_{ce}\over
 z_{ce}}\frac{ z_{et}}{ z_{at}}\right]{\epsilon_a\cdot
k_d\over  z_{ad}}{\epsilon_d\cdot\epsilon_c\over
 z_{dc}}{k_c\cdot\epsilon_b\over  z_{cb}}~.~~~\eea
Again, while keeping the weight-2 conditions for all nodes, the
linking number is decreased to $3$.

(3) The third kind of cancelation happens among diagrams (A4) and
(D3). In particular,
\bea
\begin{array}{c}
\includegraphics[width=0.9\textwidth]{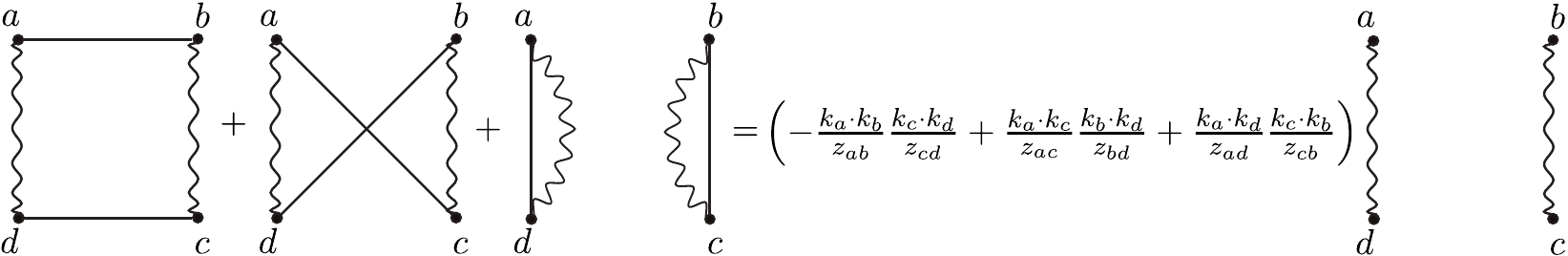}\end{array}~,~~~\label{eq:4ptDoublePole3}
\eea
where the RHS reads
\bea\left(-\frac{k_a\cdot k_b}{ z_{ab}}\frac{k_c\cdot k_d} {
z_{cd}}+\frac{k_a\cdot k_c}{ z_{ac}}\frac{k_b\cdot k_d}{ z_{bd}}
+\frac{k_a\cdot k_d}{ z_{ad}}\frac{k_c\cdot k_b}{ z_{cb}}\right)
{\epsilon_b\cdot\epsilon_c\over
 z_{bc}}{\epsilon_d\cdot\epsilon_a\over  z_{da}}~.~~~
\label{eq:4ptDoublePole3a} \eea
According to the cross-ratio identity \eqref{crossratio}, we have
the following expressions
\bea -{k_a\cdot k_b\over  z_{ab}}=(k_a\cdot k_c){ z_{ct}\over z_{ac}
z_{bt}}+(k_a\cdot k_d){ z_{dt}\over z_{ad} z_{bt}}+\Sl_{i\neq
a,b,c,d,t}(k_a\cdot k_i){ z_{it}\over z_{ai} z_{bt}}~,~~~\\
-{k_d\cdot k_b\over  z_{db}}=(k_d\cdot k_c){ z_{ct}\over z_{dc}
z_{bt}}+(k_d\cdot k_a){ z_{at}\over z_{da} z_{bt}}+\Sl_{i\neq
a,b,c,d,t}(k_d\cdot k_i){ z_{it}\over z_{di} z_{bt}}~,~~~\\
-{k_c\cdot k_b\over  z_{cb}}=(k_c\cdot k_a){ z_{at}\over z_{ca}
z_{bt}}+(k_c\cdot k_d){ z_{dt}\over z_{cd} z_{bt}}+\Sl_{i\neq
a,b,c,d,t}(k_c\cdot k_i){ z_{it}\over z_{ci} z_{bt}}~,~~~\eea
where $t\neq a,b,c,d$. Plugging these equations into
\eqref{eq:4ptDoublePole3a}, we finally obtain
\bea \Sl_{i\neq a,b,c,d,t}\left[(k_a\cdot k_i){ z_{it}\over z_{ai}
z_{bt}}\frac{k_c\cdot k_d}{ z_{cd}}+(k_d\cdot k_i){ z_{it}\over
z_{di} z_{bt}}\frac{k_a\cdot k_c}{ z_{ac}}-(k_c\cdot k_i){
z_{it}\over z_{ci} z_{bt}}\frac{k_a\cdot k_d}{
z_{ad}}\right]{\epsilon_b\cdot\epsilon_c\over
 z_{bc}}{\epsilon_d\cdot\epsilon_a\over  z_{da}}~,~~~\eea
which  decrease the linking number to $m-1=3$ while keeping the
weight-2  conditions for all nodes.

(4) The fourth type of cancelation is between diagrams (B2) and
(C6). All such contributions are collected as
\bea
\begin{array}{c}
\includegraphics[width=0.58\textwidth]{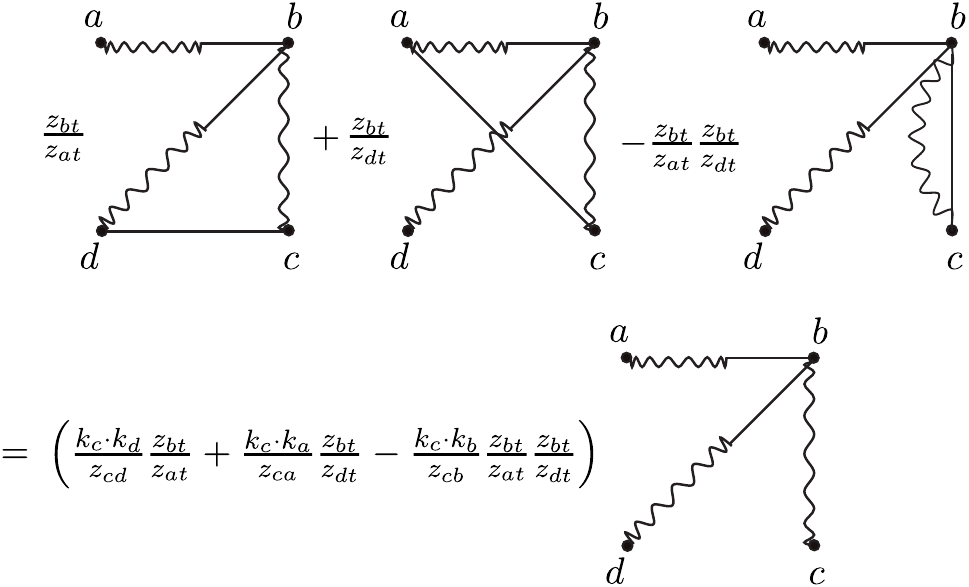}\end{array}~.~~~\label{eq:4ptDoublePole4}
\eea
The RHS is given by
\bea \frac{k_c\cdot k_d}{ z_{cd}}\frac{ z_{bt}}{
z_{at}}\left(1+{s_{ca}\over s_{cd}}{ z_{cd}\over z_{ca}}{
z_{at}\over z_{dt}}+{s_{cb}\over s_{cd}}{ z_{cd}\over z_{cb}}\frac{
z_{bt}}{ z_{dt}}\right){k_b\cdot\epsilon_a\over
z_{ba}}{\epsilon_d\cdot k_b\over
z_{db}}{\epsilon_b\cdot\epsilon_c\over z_{bc}}~.~~~\eea
Using the cross-ratio identity \eqref{crossratio}, we arrive at
\bea \frac{k_c\cdot k_d}{ z_{cd}}\frac{ z_{bt}}{
z_{at}}\left(-\Sl_{e\neq a,b,c,d,t}{s_{ce}\over s_{cd}}{ z_{cd}\over
z_{ce}}{ z_{et}\over z_{dt}}\right){k_b\cdot\epsilon_a\over
z_{ba}}{\epsilon_d\cdot k_b\over
z_{db}}{\epsilon_b\cdot\epsilon_c\over z_{bc}}~,~~~\eea
which  decreases the linking number to $m-1=3$.

\section{No higher-order poles by more general consideration}
\label{secTrunction}

In the above section, we have used explicit calculations to show the
cancelation of higher-order poles when summing over all
contributions. In this section, we will take a different approach to
study the same problem. Comparing with the previous method, this new
approach is simpler and general, which is the advantage of this
method. However, it can not present the explicit picture of how the
cancelation happens, which is an advantage of the first method.

Our starting point is to show that the reduced Pfaffian does not
contribute to the double pole. The key for this conclusion is that,
the expansion of reduced Pfaffian \eqref{eq:ExpandPf} is independent
of the gauge choice of removing the two rows and columns
\cite{Cachazo:2013hca}. Bearing this in mind, we then present the
arguments. For a given subset $A$ of $n$-elements, we can always
take the gauge choice $(\mu,\nu)$ of the reduced Pfaffian, such that
$\mu\in A$ and $\nu\not\in A$. From \eqref{order-poles-LR}, it is
known that for $\chi(A)=1$, we need subset $A$ to be given as the
union of cycles of permutation $p\in S'_n$. However, with our
special gauge choice, the cycle $W_I$ does not belong to $A$, thus
we have shown that $\chi(A)\leq 0$ for all terms in the
\eqref{eq:ExpandPf}. Since for any pole (i.e., any subset $A$), we
can always make the gauge choice to show the absence of the double
pole as above and the whole result is  independent of the gauge
choice, we have shown the absence of all possible double poles in
the reduced Pfaffian. It is worth to emphasize that in the above
argument, the independence of gauge choice for the reduced Pfaffian
 has played crucial role. However, this fact is
true based on both the gauge invariance and scattering equations, so
it is the on-shell property.

With above scenario, we can show immediately that when
\eqref{eq:ExpandPf2} appears as a factor in the CHY-integrand, it
will not contribute double pole $s_A$, where $A$ is the subset of
these $m$-particles. The argument is very easy. If we choose the
gauge $\mu,\nu\not\in A$, the reduced Pfaffian can be written as
\bea \Pf~'\Psi_n=\Pf~\Psi_m\Big(\sum
\cdots\Big)+\cdots~,~~\label{PF-kinematic}\eea
where possible double pole contribution for $s_A$ comes only from
the first term at the right handed side. However, since
$\Pf~'\Psi_n$ does not contain double pole $s_A$ and terms inside
$(\sum \cdots)$ have different structures of $\eps, k$ contractions,
consistency at both sides will immediately imply that the factor
$\Pf~\Psi_m$ will not give double poles either by providing an
overall factor $s_A$ in numerator or by decreasing the linking
number by one after using various on-shell or off-shell identities.
This claim can be used to explain the following facts,
\begin{itemize}

\item For the single trace part of
Einstein-Yang-Mills theory \cite{Cachazo:2014nsa}  given by
\bea {\cal I}_{r,s}^{\EYM}= \PT_r ( \alpha ) \Pf~\Psi_{\cal S}
 \Pf~'\Psi_n~,~~~\label{EYM-single-trace}\eea
the naive counting indicates the $\chi({\cal S})=1$.  However,
as we have argued, the factor  $\Pf~\Psi_{\cal S}$ will provide
a factor $P_{\cal S}^2$ in numerator or decrease the linking
number by one, so this double pole does not appear.

As a comparison, the double trace of gluons without gravitons in
Einstein-Yang-Mills theory has the CHY-integrand
\cite{Cachazo:2014nsa}
\bea {\cal I}_{r+s}^{\EYM}= s_{r}{\rm PT}_r (\alpha) {\rm
PT}_s(\beta) {\rm Pf}~'\Psi_n~.~~~\label{EYM-double-trace}\eea
The double cycle ${\rm PT}_r (\alpha) {\rm PT}_s(\beta)$ will
generate manifest double pole $s_r^2$ when one integrates
$z_i$'s, thus the explicit kinematic factor $s_{r}$ is needed to
make it to be physical amplitude.

\item For Yang-Mills-Scalar theory  with $q$ scalars and $r=n-q$ gluons
the CHY-integrand is given by \cite{Cachazo:2014nsa}
\bea {\cal I}^{\YMs}= {\rm PT}_n (\alpha) \left(  {\rm PT}_q
(\alpha)~ {\rm Pf}~\Psi_{r}\right)~.~~~\label{YM-S-trace}\eea
The  naive double pole $s_q^2$ from $z$-integration will be
canceled by the kinematic numerator factor $s_q^2$ provided by
${\rm Pf}~\Psi_{r}$ (the part with effectively reduced linking
number will not give double pole after $z$-integration). Similar
argument holds for more general CHY-integrand with $q$ scalars,
$r$ gluons and $s=n-q-r$ gravitons,
\bea {\cal I}= \Big({\rm PT}_{q+r} (\alpha) ~{\rm Pf}
~\Psi_{s}\Big)\times \Big(  {\rm PT}_q (\beta)~ {\rm Pf}~
\Psi_{r+s}\Big)~.~~~\label{YM-S-GR-trace}\eea
So naive double poles of $P_s^2$ and  $P_{s+r}^2$ will not
appear.

\end{itemize}
%

\subsection{Dimensional reduction to EYM theory}

Argument given in \eqref{PF-kinematic} has shown that ${\rm
Pf}~\Psi_A$ will contribute double pole of  $s_A$. However, it is
not obvious that the double poles $s_{B\subset A}$ in ${\rm
Pf}~\Psi_m$ (for example, the CHY -integrand
\eqref{EYM-single-trace}) will not appear. To understand this point,
we can use the technique of dimensional reduction.

To demonstrate the method, let us focus on the single trace part of
Einstein-Yang-Mills theory given in \eqref{EYM-single-trace}. We
start from gravity CHY-integrand $\Pf~'
\Psi_{n}(k_i,\epsilon_i,z_i)~\Pf~' \Psi_{n}(k_i,\W \epsilon_i,z_i)$,
which gives result containing only single poles for all allowed
physical configurations. Now we divide $n$ particles into two
subsets, $1,2,\dots, m\in\{\mathfrak{g}\}$ and $m+1,m+2,\ldots,
n\in\{\mathfrak{h}\}$ and  assign the particular physical
configurations as follows. Firstly, all momenta in
$(D+d)$-dimensions are split into $D$-dimensional part and
$d$-dimensional part as
\bea \{(k_1,\eta),(k_2,-\eta),(k_i,0)\}~~~,~~~i=3,\ldots,n,
~~~~k_i\in R^{1,D-1}~~~,~~~\eta\in
R^{d}~,~~~\label{1Trace-trun-1}\eea
where on-shell conditions require
\bea \eta^2=0~~~,~~~k_i^2=0~~~,~~~i=1,2,\ldots,n~,~~~\sum_{i=1}^n
k_i=0~.~~~\label{1Trace-trun-2}\eea
Secondly, the polarization vectors are taken as
\bea
\{(0,\W\epsilon_1)~,~(0,\W\epsilon_2)~,~(0,\W\epsilon_i)~,~(\W\epsilon_j,0)\}~~~,~~~i=3,4,\ldots,m~,~~~j=m+1,m+2,\ldots,n~,~
~~\label{1Trace-trun-3}\eea
which satisfy
\bea \eta\cdot\W\epsilon_1=\eta\cdot
\W\epsilon_2=0~~,~~\W\epsilon_i=\W\epsilon~,~~~i=3,\ldots,m~~,~~\W\epsilon^2=0~~,~~\W\epsilon_1
\cdot \W\epsilon=0~~, ~~\W\eps_j\cdot
k_j=0~~,~~j=m+1,\ldots,n~.~~~\label{1Trace-trun-4}\eea
This condition can always be  achieved when $d$ is large enough. It
is obvious that when we do the dimensional reduction from $(D+d)$ to
$D$, polarization assignment in \eqref{1Trace-trun-3} means that,
particles $\{1,\ldots,m\}$ will become the gluons while particles
$\{m+1,...,n\}$ will remain to be gravitons. Having imposed these
conditions, we can see,
\begin{itemize}

\item The scattering equation in the full $(D+d)$-dimensions also implies
the scattering equations in $D$-dimensions since all $K_i\cdot
K_j=k_i\cdot k_j$.

\item The $C_{ii}$ for gluon subset are given by
\bea
C_{11}=C_{22}=0~~~,~~~C_{ii}=\W\epsilon\cdot\eta{z_{12}\over
z_{1i}z_{i2}}~~~,~~~i=2,3,\ldots,m~,~~~\label{1Trace-trun-7}
\eea
where we have chosen the gauge $t=2$ for $i=3,\ldots,m$. The
$C_{kk}$ for graviton subset are given as
\bea C_{kk}=\Sl_{j\neq k}\W\epsilon_k\cdot k_j{z_{jt}\over
z_{jk}z_{kt}}~~~,~~~k=m+1,\ldots,n~,~~~\label{1Trace-trun-8}\eea
which are nothing but those in Pfaffian of gravitons in
$D$-dimensions.
\end{itemize}

Now we evaluate the $(D+d)$-dimensional reduced Pfaffian $\Pf~'
\Psi_{n}(k_i,\W \epsilon_i,z_i)$ by chosen the gauge $(1,2)$. For
this choice, the allowed permutations will be the following cycle
structures,
\bea (1\a_1 2) (\a_2)\cdots(\a_m) C_{j_1j_1}\cdots C_{j_t
j_t}~,~~~\label{1Trace-trun-9}\eea
and we consider these cycles one by one as,
    \begin{itemize}

    \item For $W$-cycle, the numerator is
    \bea \eps_1\cdot U_{\a_1(1)} U_{\a_2(2)}\cdots  \cdot\eps_2
    ~.~~~\label{1Trace-trun-10}\eea
    Now using $U_i=k_i \eps_i-\eps_i k_i$ and imposing
    conditions \eref{1Trace-trun-3}, \eref{1Trace-trun-4}, we
    see that for all subsets $\a_1 \subset \{3,4,\ldots,n\}$,
    the contraction is zero. So the only non-zero contribution
from $W$-cycle is when $\a_1=\emptyset$ with factor
\bea {\W\eps_1\cdot \W\eps_2\over
\spaa{12}}~.~~~\label{1Trace-trun-11}\eea

    \item For $U$-cycle with at least two elements, if $i\in \{3,\ldots,r\}$ inside an $U$-cycle,
     the combination
    \bea U_k \cdot (k_i \W\eps_i-\W\eps_i k_i)\cdot U_t
    ~~~\label{1Trace-trun-12}\eea
    will be zero, since by our reduction conditions, for any
    $k\in \{3,4,\ldots,n\}$ we will always have
    \bea U_k\cdot \eps_i=0~.~~~\label{1Trace-trun-13} \eea
    In other words, any $i\in \{3,4,\ldots,r\}$ can not be
    inside an $U$-cycle with at least two elements.

    \end{itemize}
With above discussions, we see that non-zero contributions are
    \bea \Pf~'\Psi_{n}(k_i,\W
    \epsilon_i,z_i)\to {\W\eps_1\cdot \W \eps_2\over \spaa{12}}
    C_{33}\cdots C_{mm} \left\{\sum\cdots \right\}={\W\eps_1\cdot \W
    \eps_2\over \spaa{12}} C_{33}\cdots C_{mm} {\rm Pf}~\Psi_{\cal G}
    ~.~~~\label{1Trace-trun-14}\eea
Result \eqref{1Trace-trun-14} is not the form
\eqref{EYM-single-trace} we are looking for. To reach that, we must
use \eqref{1Trace-trun-7} and the insertion relation
(\ref{Building-II-1}). Thus
\bea{\W\eps_1\cdot \W \eps_2\over (12)} C_{33}\cdots
C_{rr}=\sum_{\a}
(\W\epsilon_1\cdot\W\epsilon_2)(\W\epsilon\cdot\eta)^{m-2}{\rm PT}
(1\alpha(3,\cdots,m)2)~.~~~\label{1Trace-trun-14-1} \eea

Combining \eqref{1Trace-trun-14} and \eqref{1Trace-trun-14-1}, we
see that the $\Pf~'\Psi_{n}(k_i,\W \epsilon_i,z_i)$ indeed has been
reduced to the sum of the form ${\rm PT} (1\alpha(3,\cdots,m)2){\rm
Pf}~\Psi_{\cal G}$.

Having finished the part $\Pf~'\Psi_{n}(k_i,\W \epsilon_i,z_i)$, we
are left with the part $\Pf~'\Psi_{n}(k_i, \epsilon_i,z_i)$, which
is in $(D+d)$-dimension. To reduce to $D$-dimension, we must impose
proper choice of polarization vectors $\eps_i^{(D+d)}$. It is easy
to see that the choice $\eps_i^{(D+d)}=(\eps_i,0)$ will do the job.

Putting all together we see that, starting from $(D+d)$-dimensional
gravity theory, we do able to reduce to single trace part of EYM
theory with CHY-integrand \eqref{EYM-single-trace}\footnote{If
taking the graviton subset to be empty, we have reduced the gravity
theory to Yang-Mills theory.}. Since the gravity theory does not
contain any double poles, so is \eqref{EYM-single-trace}. This
finishes our general proof.

\subsection{Dimensional reduction to  $(\Pf~'A_n)^2$}

In effective theories, such as non-linear sigma model and
Dirac-Born-Infeld theory, we also encounter $(\Pf~'A_n)^2$. This
$(\Pf~'A_n)^2$ can also be obtained from $\Pf~'\Psi$ by taking
appropriate dimensional reduction. Specifically, we impose momenta
and polarization vectors in $(d+d+d)$-dimensions as follows,
\bea K_a= (k_a;0; 0)~~~,~~~\WH \eps_{a\neq \a,\b}=(0;
\eps_a;0)~~~,~~~\WH
\eps_{i}=(0;0;\eps_i)~~~,~~~i=\a,\b~~~,~~~a=1,2,\ldots,n~,~~~\label{CHY-trun-A2-3}\eea
where $\a,\b$ are the gauge choice for reduced Pfaffian. With this
assignment $K_a\cdot \WH \eps_b=0$, so transverse condition of
polarization vector has kept and the $C$-block of matrix $\Psi$ is
zero. Thus we have
\bea {\rm Pf}~' \Psi^{(d+d+d)}={\rm Pf}~' A^{(d+d+d)}~~ {\rm
Pf}~B^{(d+d+d)}~.~~~\label{CHY-trun-A2-4}\eea
Furthermore, with the choice in \eqref{CHY-trun-A2-3}, we see two
facts. First, the ${\rm Pf}~' A$ in $(d+d+d)$-dimension is in fact
in $d$-dimension. Second, $\WH \eps_{i}\cdot \WH \eps_a=0$ when
$i=\a,\b$ and $a\neq \a,\b$, thus we have
\bea {\rm Pf}~B= {(-)^{\a+\b} \eps_\a\cdot \eps_\b\over
z_{\a\b}}{\rm Pf}~ B^{\a\b}_{\a\b}\sim \ell^2 k_\a\cdot k_\b  ~{\rm
Pf}~' A|_{\eps_a\to \ell k_a}~,~~~\label{CHY-trun-A2-5}\eea
where in the last step, we have set $\eps_i=\ell k_i$ for
$i=1,\ldots,n$. Putting all together, we see that up to factor
$\ell^2 k_\a\cdot k_\b$, we do dimensionally reduce the reduced
Pfaffian to $(\Pf~'A_n)^2$.

There is one obvious generalization.  Instead of just two $\a,\b$,
we divide all $n$-particles into $m$ groups, and polarization
vectors of each group belongs to independent subspace. Then we can
take $\eps_a\sim k_a$, so
\bea {\rm Pf}~ B \to \prod_{i=1}^m {\rm Pf}
~A_i~.~~~\label{CHY-trun-A2-6}\eea
%

\section{Conclusions}
\label{secConclusion}

In this paper, we systematically discuss the cancelation of
higher-order poles  in CHY-formula. By expanding the cycles of
(reduced) Pfaffian into pieces we established a diagrammatic
representations. Grouping diagrams appropriately and applying
cross-ratio identity, we show that the linking number for a pole
$s_{A}$ receives a value of $|A|-1$ from the Pfaffian. This means
there is no any higher-order poles in Yang-Mills theory and gravity.
We then developed the dimensional reduction procedures, by which
integrands of other theories can be produced from gravity theory.
Thus higher-order poles will not exist in these theories by the
consistent reduction.

Inspired by results in this paper, there are several interesting
questions worth to investigate. The first thing is that although
with explicit examples of two, three, four points, we have shown the
pattern how the explicit cancelation of double poles work, writing
down the general explicit argument is still welcome.

Another thing is that, in papers
\cite{Cachazo:2014xea,Cachazo:2016njl}, CHY-integrands for various
field theories have been proposed through various techniques, such
as compactifying, generalized dimensional reduction, generalizing,
squeezing and extension from soft limit, etc. Starting from a
physical meaningful mother theory\footnote{Here the physical
meaningful theory means the corresponding tree-level amplitudes
possess only single pole, correct factorization and soft limits,
etc.}, some techniques guarantee a physical meaningful daughter
theories at the end, such as the compactifying and generalized
dimensional reduction. This is exactly the aspect we are using in
this paper. However, some techniques, such as squeezing and
extension from soft limits, are not so obvious to produce physical
meaningful daughter theories at the end. Thus it is definitely
important to study these techniques further and to see if all these
different techniques can be unified from a single picture.
Furthermore, finding the algorithm to read out the daughter theory
(i.e., its field contents and Lagrangian) from the known mother
theory in various construction techniques is also an interesting
question.

\section*{Acknowledgments}

We would like to thank useful discussions with Fei Teng and Chih-Hao
Fu. BF is supported by Qiu-Shi Funding and the National Natural
Science Foundation of China (NSFC) with Grant No.11135006,
No.11125523 and No.11575156. YD would like to acknowledge NSFC under
Grant Nos.11105118, 111547310, as well as the support from 351 program of Wuhan
University. RH would like to acknowledge the supporting from
NSFC No.11575156 and the Chinese Postdoctoral Administrative
Committee.

\appendix

\section{The on-shell and off-shell identities of CHY-integrands}
\label{appendixIdentity}

In the decomposition of reduced Pfaffian as sum of Parke-Taylor
factors, we have taken advantages of many non-trivial relations
between rational functions of $z_i$'s. Some relations are valid at
the algebraic level, and we call them off-shell relations. The
others are valid only when $z_i$ takes values of the solutions of scattering equations and we
call them on-shell ones. The most important one of the latter case
is the cross-ratio identities \eref{crossratio} , derived from the original scattering
equations. Any others can be derived from them. For the off-shell identities, we borrow the name from
amplitude relations and have (recall the notation \eref{notation})
\begin{itemize}
  \item The Schouten identity,
  \bea {[a~b]\over [a~c][c~b]}={[a~d]\over
  [a~c][c~d]}+{[d~b]\over
  [d~c][c~b]}~,~~~\label{schoutenIden}\eea
  \item The $U(1)$-decoupling relation,
  \bea \sum_{\shuffle}{1\over \langle
  a_1,\{a_2,\ldots,a_{n}\}\shuffle\{b\}\rangle}=0~,~~~\label{u1Iden}\eea
  \item The KK-relation,
  \bea {1\over \langle
  a_1,\alpha,a_n,\beta\rangle}=(-)^{n_{\beta}}\sum_{\shuffle}{1\over
  \langle
  a_1,\alpha\shuffle\beta^T,a_n\rangle}~,~~~\label{KKIden}\eea
  where $n_\beta$ is the number of elements in set $\beta$, and
  $\beta^T$ is the reverse of set $\beta$.
\end{itemize}
The Schouden identity is trivial, by understanding that
\bea [a~b][c~d]=[a~c][b~d]+[a~d][c~b]~.~~~\nonumber\eea
\noindent{\bf The proof of $U(1)$-relation:} to prove the
$U(1)$-decoupling relation, let us start from the Schouten identity
\bea {[a_1,a_{n}]\over [a_1,b][b,a_{n}]}={[a_1,a_{n-1}]\over
[a_1,b][b,a_{n-1}]}+{[a_{n-1},a_{n}]\over
[a_{n-1},b][b,a_{n}]}~.~~~\eea
Repeatedly using the Schouten identity
\bea {[a_1,a_k]\over [a_1,b][b,a_k]}={[a_1,a_{k-1}]\over
[a_1,b][b,a_{k-1}]}+{[a_{k-1},a_k]\over [a_{k-1},b][b,a_k]}~~~~\eea
until $k=2$, we get
\bea{[a_1,a_{n}]\over
[a_1,b][b,a_{n}]}=\sum_{k=2}^{n}{[a_{k-1},a_k]\over
[a_{k-1},b][b,a_k]}~.~~~\label{temp-2}\eea
Then
\bea {1\over \langle a_1,a_2,\ldots,a_{n},b\rangle}&=&-{1\over
\langle a_1,a_2,\ldots, a_{n}\rangle}\left({[a_1,a_{n}]\over
[a_1,b][b,a_{n}]}\right)\nonumber\\
&=&-\sum_{k=2}^{n}{1\over \langle a_1,a_2,\ldots,
a_{n}\rangle}{[a_{k-1},a_k]\over
[a_{k-1},b][b,a_k]}~,~~~\label{U1Iden2}\eea
which is nothing but the $U(1)$-relation after canceling the
numerator $[a_{k-1},a_k]$ with its corresponding factor in the
$\langle a_1,\ldots, a_{n}\rangle$.

\noindent {\bf The proof of the KK-relation:} the KK-relation can be
proven by induction\footnote{Similar discussions can be found in \cite{Kol:2014yua}.}. For $n_\beta=1$, it is the $U(1)$-relation,
which has already been proven. Assuming (\ref{KKIden}) is valid for
$\beta=\{\beta_1,\beta_2,\ldots,\beta_m\}$, when $n_\beta=m+1$, we
have
\bea{1\over \langle
a_1,\alpha,a_n,\beta_1,\beta_2,\ldots,\beta_m,\beta_{m+1}\rangle}&=&{1\over
\langle
a_1,\alpha,a_n,\beta_1,\beta_2,\ldots,\beta_m\rangle}{[\beta_m,a_1]\over[\beta_m,\beta_{m+1}][\beta_{m+1},a_1]
}\nonumber\\
&=&\sum_{\shuffle}(-)^{m}{1\over \langle
a_1,\alpha\shuffle\{\beta_m,\beta_{m-1},\ldots,\beta_1\},a_n\rangle}{[\beta_m,a_1]\over[\beta_m,\beta_{m+1}][\beta_{m+1},a_1]
}~.~~~\eea
The second line can be rewritten as
\bea &&\sum_{s=1}^{n_\alpha}\sum_{\shuffle}(-)^{m}{1\over [
a_1,\alpha_1,\ldots,\alpha_s,\beta_m][\beta_m,\{\alpha_{s+1},\ldots,\alpha_{n_\alpha}\}\shuffle\{\beta_{m-1},\ldots,\beta_1\},a_n][a_n,a_1]}{[\beta_m,a_1]\over[\beta_m,\beta_{m+1}][\beta_{m+1},a_1]
}\nonumber\\
&&=\sum_{s=1}^{n_\alpha}\sum_{\shuffle}(-)^{m}{[\beta_m,a_1]\over
\langle
a_1,\alpha_1,\ldots,\alpha_s,\beta_m,\beta_{m+1}\rangle[\beta_m,\{\alpha_{s+1},\ldots,\alpha_{n_\alpha}\}\shuffle\{\beta_{m-1},\ldots,\beta_1\},a_n][a_n,a_1]}\nonumber\\
&&=\sum_{s=1}^{n_\alpha}\sum_{\shuffle,\shuffle_1}(-)^{m+1}{[\beta_m,a_1]\over
\langle
a_1,\{\alpha_1,\ldots,\alpha_s\}\shuffle_1\{\beta_{m+1}\},\beta_m\rangle[\beta_m,\{\alpha_{s+1},\ldots,\alpha_{n_\alpha}\}\shuffle\{\beta_{m-1},\ldots,\beta_1\},a_n][a_n,a_1]}\nonumber\\
&&=\sum_{\shuffle}(-)^{m+1}{1\over \langle
a_1,\alpha\shuffle\{\beta_{m+1},\beta_m,\ldots,\beta_1\},a_n\rangle}=\sum_{\shuffle}(-)^{n_\beta}{1\over
\langle a_1,\alpha\shuffle\beta^{T},a_n\rangle}~,~~~\eea
where in the third line we have used the $U(1)$-relation for
$1/\langle a_1,\alpha_1,\ldots,\alpha_s,\beta_m,\beta_{m+1}\rangle
$. This ends the induction proof.

Some implications can be deduced from the $U(1)$-relation
(\ref{u1Iden}) and KK-relation (\ref{KKIden}). From the
$U(1)$-relation with the expression (\ref{U1Iden2}),
\bea {1\over [
a_1,a_2,\ldots,a_{n}][a_{n},b][b,a_1]}&=&-\sum_{k=2}^{n}{1\over
[a_n,a_1][ a_1,a_2,\ldots, a_{n}]}{[a_{k-1},a_k]\over
[a_{k-1},b][b,a_k]}~,~~~\eea
we immediately get the so called insertion relation,
\bea {1\over [ a_1,a_2,\ldots,a_{n}]}{[a_1,a_n]\over
[a_{1},b][b,a_n]}&=&\sum_{i=1}^{n-1}{1\over [
a_1,\ldots,a_i,b,a_{i+1},\ldots,
a_{n}]}~,~~~\label{Building-II-1}\eea
where we have inserted the node $b$ between $a_1$ and $a_n$. From
the KK-relation (\ref{KKIden}), we get the so called  open-up
relation,
\bea {[a_1,a_n]\over \langle
  a_1,\alpha,a_n,\beta\rangle}=(-)^{n_{\beta}+1}\sum_{\shuffle}{1\over
  [
  a_1,\alpha\shuffle\beta^T,a_n]}~,~~~\label{openIden}\eea
which opens a closed cycle $\langle a_1,\alpha,a_n,\beta\rangle$ to
a sum of open cycles. This relation can be trivially seen by
applying KK-relation (\ref{KKIden}) for the denominator. We can diagrammatically
abbreviate it as,
\bea
\begin{array}{c}
  \includegraphics[width=3in]{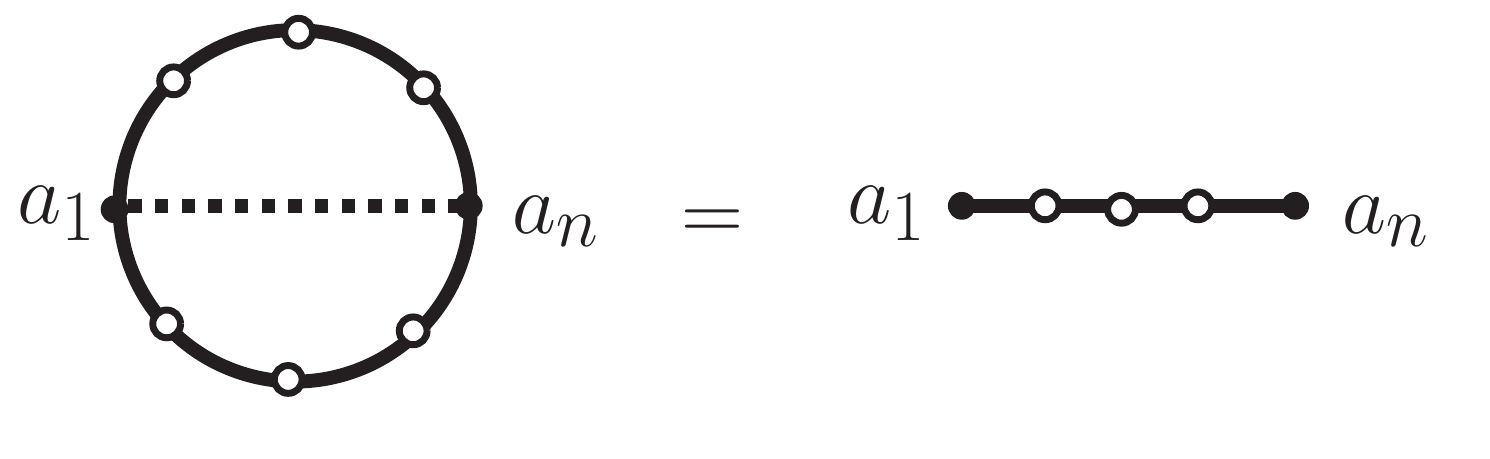}
\end{array}~,~~~\label{FigOpenup}
\eea
where a line with white dots means there are other $z_i$'s locating
along the line, with its explicit definition in (\ref{openIden}).

\noindent {\bf A sketch of expanding into PT-factors:} having
presented the off-shell and on-shell identities, now we show how to
use them to simplify the CHY-integrand to the PT-factors, which are
easily evaluated by integration rule method
\cite{Baadsgaard:2015voa,Baadsgaard:2015ifa,Baadsgaard:2015hia,Huang:2016zzb}
 without referring to the
scattering equations. This algorithm has been laid out in
\cite{Bjerrum-Bohr:2016axv}, but here we provide an alternative
understanding. It is trivial to see that any weight-2 CHY-integrand
can be written as product of a PT-factor with $n$ nodes and
cross-ratio factors such as ${[a_i~a_j][a_k~a_\ell]\over
[a_i~a_k][a_j~a_\ell]}$. Thus by showing the reduction of one
cross-ratio factor is suffice to explain any situations. Let us
focus on the following CHY-integrand, given as
\bea {1\over \langle
a_1,a_2,\ldots,a_n\rangle}{[a_i~a_j][a_k~a_\ell]\over
[a_i~a_k][a_j~a_\ell]}=\left({[a_i~a_j]\over \langle
a_1,a_2,\ldots,a_n\rangle}\right){[a_k~a_\ell]\over
[a_i~a_k][a_j~a_\ell]}~.~~~\label{termCHY}\eea
Applying (\ref{openIden}) to the expression in the bracket, we will
get two possible results for (\ref{termCHY}), as
\bea
\begin{array}{c}
  \includegraphics[width=5in]{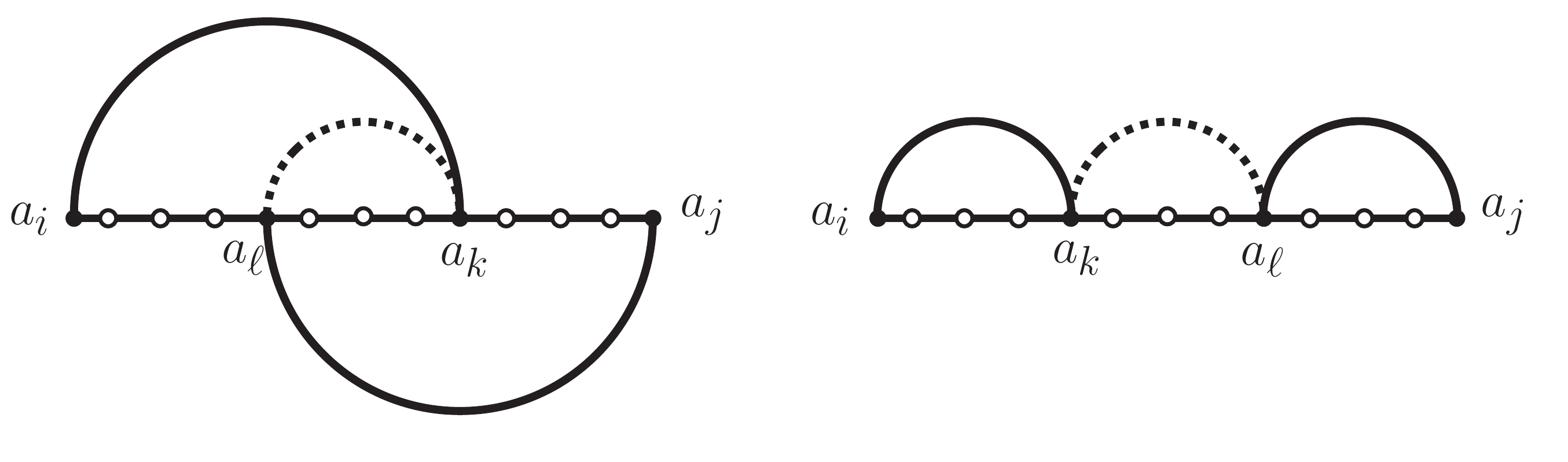}
\end{array}~~~~\label{FigtermCHY1}
\eea
where the expression in the bracket leads to the line with white
dots from $z_i$ to $z_j$, and the other factors denoted by half
circles. For the first situation in (\ref{FigtermCHY1}), we can
again apply (\ref{FigOpenup}) to the up-half plane, which ends up
with
\bea
\begin{array}{c}
  \includegraphics[width=2in]{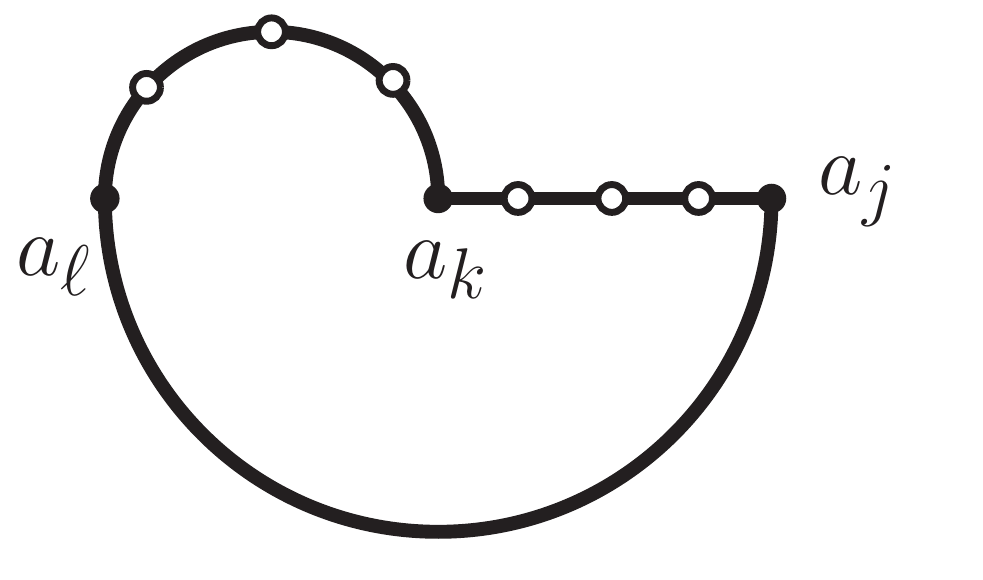}
\end{array}~,~~~\label{FigtermCHY2}
\eea
which is a PT-factor. For the second situation in
(\ref{FigtermCHY1}), we shall use the cross-ratio identity
\bea -1=\sum_{{i'\in A/\{k\}\atop j'\in A^c/\{\ell\}}}
{s_{i'j'}\over s_A}{[a_{i'}~a_k][a_{j'}~a_\ell]\over
[a_{i'}~a_{j'}][a_k~a_\ell]}~,~~~\label{termCHYcross}\eea
where we choose set $A$ to be collection of $z_i$'s in the left-most
cycle, so $j'$ runs over white dots in between $a_k, a_\ell$ or
those in between $a_\ell,a_j$. The factor $[a_k~a_\ell]$ in
denominator cancels the dashed line, so after multiplying
(\ref{termCHYcross}) to the second figure of (\ref{FigtermCHY1}), we
get the following contributions depending on the location of $a_{j'}$,
\bea
\begin{array}{c}
  \includegraphics[width=6in]{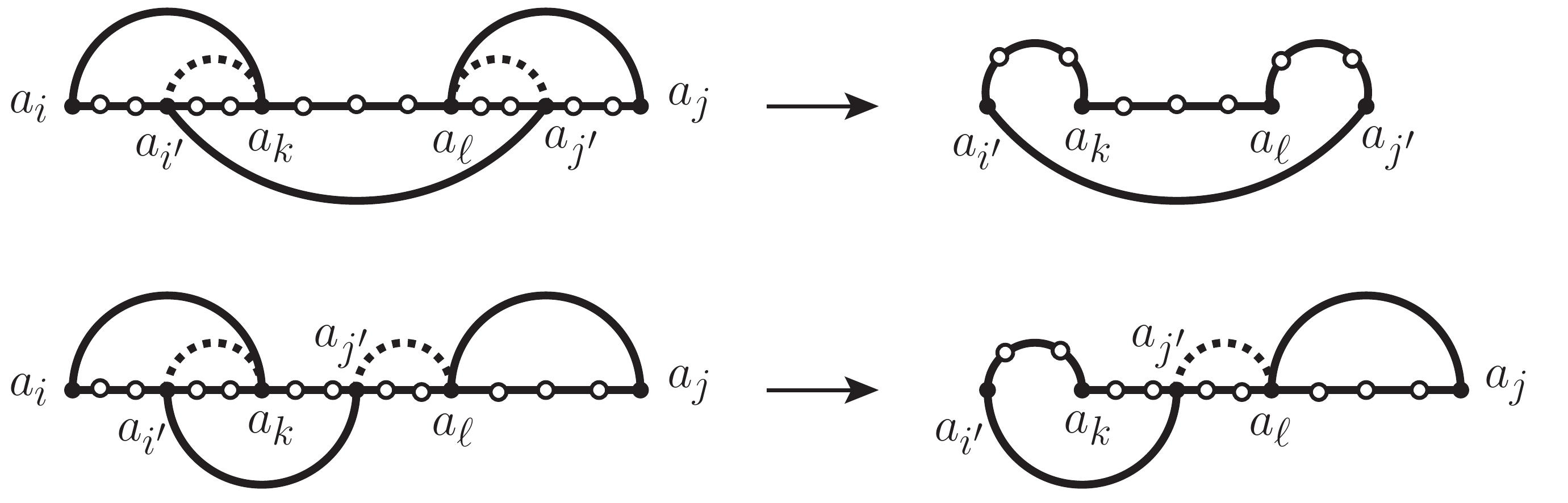}
\end{array}~.~~~\label{FigtermCHY3}
\eea
The result in the first line is already PT-factor, while the result
in the second line has the same structure as the second figure in
(\ref{FigtermCHY1}), but with fewer $z_i$'s in between
$a_{j'},a_\ell$. Recursively applying cross-ratio identity, we will
end up with the situation where there is no $z_i$ in between
$a_{j'}, a_\ell$, hence the dashed line is canceled and we get two
disjoint cycles. In such case, we can apply cross-ratio identity
again as
\bea
\begin{array}{c}
  \includegraphics[width=4in]{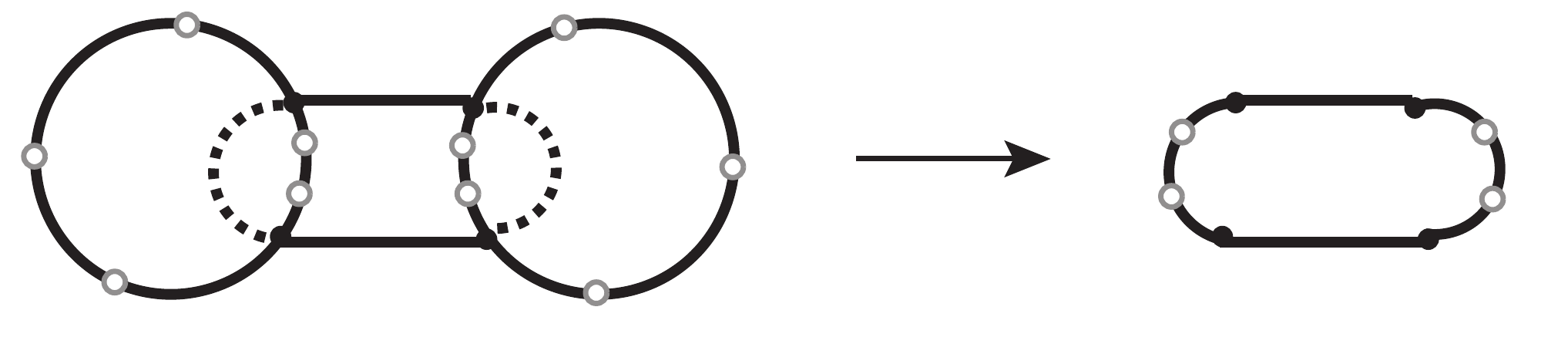}
\end{array}~,~~~\label{FigtermCHY4}
\eea
where for the cross-ratio identity (\ref{termCHYcross}) we have
chosen $a_{i'},a_{k}$ in one cycle and $a_{j'},a_{\ell}$ in the
other cycle. Then applying the open-up identity (\ref{FigOpenup}) in
both sides, we get the desired result.


\bibliographystyle{JHEP}
\bibliography{Dpole}

\end{document}